\documentstyle[preprint,eqsecnum,aps]{revtex}
\begin{document}
\title{The Schwinger SU(3) Construction-II: Relations between
Heisenberg-Weyl and SU(3) Coherent States}
\author{S. Chaturvedi\thanks{email: scsp@uohyd.ernet.in}}
\address{ School  of Physics, University of Hyderabad, Hyderabad 500046,
India}
\author{N. Mukunda\thanks{email: nmukunda@cts.iisc.ernet.in} \thanks{Honorary 
Professor, Jawaharlal Nehru Centre for
Advanced Scientific Research, Jakkur, Bangalore 560064}}
\address{Centre for Theoretical Studies, 
Indian Institute of Science,
Bangalore 560012, India}

\date{\today}
\maketitle
\begin{abstract}
The Schwinger oscillator operator representation of $SU(3)$, studied in a 
previous paper from the representation theory point of view, is analysed to 
discuss the intimate relationships between standard oscillator coherent state 
systems and systems of $SU(3)$ coherent states. Both $SU(3)$ standard coherent 
states, based on choice of highest weight vector as fiducial vector, and
certain other specific systems of generalised coherent states, are found to 
be relevant. A complete analysis is presented, covering all the oscillator 
coherent states without exception, and amounting to $SU(3)$ harmonic analysis
of these states.  
\end{abstract}

\section{Introduction}

In a previous paper\cite{1} we have presented an analysis of the reducible 
unitary representation(UR) of $SU(3)$ that is obtained by a 
generalisation of the well-known Schwinger oscillator operator 
construction in the case of $SU(2)$\cite{2}.  This construction, based
on six independent pairs of oscillator operators, is a minimal one 
in the sense that all unitary irreducible representations (UIR) of 
$SU(3)$ are obtained without exception.  However in contrast to the 
$SU(2)$ case there is an unavoidable multiplicity in that each  UIR
occurs a denumerably infinite number of times.  A systematic way to 
handle this multiplicity, based on the use of the non compact group
$Sp(2,R)$, has been developed; its salient features are recapitulated
in the next Section.

The aim of the present paper is to extend this study and 
discuss various properties of coherent states in this framework.  
The use of oscillator operators automatically brings in the
Heisenberg-Weyl (H-W)  group with a dimension appropriate to the number
of independent oscillators or degrees of freedom.  And it is 
indeed in the context of this group that the standard 
coherent states in quantum mechanics were originally defined
and applied to a very large number of problems\cite{3}.  On the other
hand, the basic kinematic relations for any system of independent
oscillator operators have a well-defined covariance group associated 
with them - a group of linear inhomogeneous transformations on 
the oscillator operators which leave their commutation relations 
invariant.  The homogeneous part of this covariance group is the 
metaplectic group of appropriate dimension, containing a unitary 
group as its maximal compact subgroup.  Thus for $n$ oscillators or
$n$  canonical pairs of degrees of freedom, we encounter the 
groups $Mp(2n)$, $U(n)$ and $SU(n)$, and certain of their UR's,
in a natural way \cite{4}.

Now the original concept of coherent states has  been generalised 
from the H-W case to a general Lie group, and it consists 
of the orbit of a chosen fiducial vector under group action in any UIR
of the group\cite{5}.  The usual coherent states arise by the action of the
elements of the H-W group on the Fock vacuum.  Given 
all this, it is natural and to be expected that via the Schwinger 
type construction we have an intricate interplay between the 
familiar H-W coherent states, and certain systems of
coherent states associated with the groups $Mp(2n), U(n)$ and
$SU(n)$.  

In passing we may also mention that with this generalisation, 
even for the H-W group we have not only the originally
defined coherent states, which may be called Standard Coherent
States (SCS), but other systems of generalised coherent states (GCS)\cite{6}.
These are based on choices of states other than the Fock vacuum 
as the fiducial state.  Similarly, for the unitary group $SU(n)$, 
within any given UIR the SCS are obtained when the highest weight 
state is used as the fiducial state, while for other choices we
have systems of GCS\cite{7}.  It is therefore of interest to see how these
various systems of coherent states for different groups get 
interconnected via the Schwinger construction.  This is the main 
aim of the present work, in the particular case of the H-W group for 
six oscillators, and $SU(3)$.

A brief outline of this work is as follows. Our earlier work \cite{1} 
has shown how in a natural manner we can identify and isolate a subspace 
${\cal H}_0$ carrying a complete and multiplicity-free UR of $SU(3)$ 
( a `Generating Representation' for $SU(3)$), within the full Schwinger 
representation characterised by infinite multiplicity. As this decomposition, 
in which the compact generator $J_0$ of $Sp(2,R)$ plays a crucial role, 
provides the starting point of the present work, to set the notation and to 
make the paper reasonably self-contained, we briefly recapitulate the 
relevant details of \cite{1} in Section II. In Section III, we recall the 
largely familiar interconnections between H-W and $U(1)$ and $SU(2)$ coherent 
states, to highlight some special features of the Klauder resolution of the 
identity and its modifications. This helps set the stage for a unified 
analysis of the relations between the appropriate H-W SCS and $SU(3)$ SCS 
and GCS carried out in detail in Sections IV, V, and VI. Section IV contains 
a detailed classification of the orbits of H-W SCS under $SU(3)$ action; we 
identify both generic orbits of maximal dimension, and non generic lower 
order ones. The rest of Section IV carries out the $SU(3)$ harmonic analysis 
of generic orbits lying in the subspace ${\cal H}_0$. In Section V we examine
the remaining generic orbits, lying in subspaces ${\cal H}_{\kappa}$ which 
are generalisations of ${\cal H}_0$ and are labelled by a complex parameter 
$\kappa$. Some calculational  details pertaining to this Section are put 
together in an appendix. Section VI contains an analysis of the $SU(3)$ 
content of a family of H-W SCS belonging to a non generic orbit under $SU(3)$ 
action. Some concluding remarks are presented in Section VII.    

\section{Review of Schwinger construction for $SU(3)$}

This construction uses six independent sets of oscillator creation 
and annihilation operators $\hat{a}^{\dag}_j, \hat{b}^{\dag}_j, 
\hat{a}_j, \hat{b}_j, j=1,2,3$, among which the only non vanishing 
commutators are
     \begin{eqnarray}
     [\hat{a}_j, \hat{a}_k^{\dag}] = [\hat{b}_j, \hat{b}_k^{\dag}] =
     \delta_{jk},\;j,k=1,2,3 .
     \end{eqnarray}

The Hilbert space ${\cal H}$ carrying an irreducible representation of 
these operators is the tensor product ${\cal H}={\cal H}^{(a)} \times
{\cal H}^{(b)}$, where ${\cal H}^{(a)}$ and ${\cal H}^{(b)}$ are the 
individual Hilbert spaces carrying irreducible representations of the 
independent sets $\hat{a}_j, \hat{a}_j^{\dag}$ and 
$\hat{b}_j, \hat{b}_j^{\dag}$ respectively.  The Schwinger UR of
$SU(3)$ acts on ${\cal H}$, and its hermitian generators are\cite{1}
     \begin{eqnarray}
     Q_{\alpha} &=& Q_{\alpha}^{(a)} + Q_{\alpha}^{(b)},\nonumber\\
     Q^{(a)}_{\alpha}&=& \frac{1}{2} \hat{a}^{\dag} \lambda_{\alpha}
     \hat{a},\;Q^{(b)}_{\alpha} = -\frac{1}{2}
     \hat{b}^{\dag} \lambda_{\alpha}^*\;\hat{b},\;
     \alpha=1,2,\ldots, 8.
     \label{2}
     \end{eqnarray}

\noindent
Here $\frac{1}{2}\;\lambda_{\alpha}$ are the eight hermitian 
traceless $3\times 3$ matrices generating the defining 
UIR $(1,0)$ of $SU(3)$\cite{8}.(For ease in writing, the UIR's of 
$SU(3)$ will be denoted by $(p,q)$ where $p,q=0,1,2,\ldots,$ 
independently, instead of the more elaborate notation $D^{(p,q)}$).

The independent mutually commuting generators $Q_{\alpha}^{(a)}, 
Q^{(b)}_{\alpha}$ lead to specific multiplicity-free UR's 
${\cal U}^{(a)}(A), {\cal U}^{(b)}(A)$ of $SU(3)$ on
${\cal H}^{(a)}, {\cal H}^{(b)}$ respectively.  Here $A$
is a general matrix in the UIR $(1,0)$.  The UR 
${\cal U}^{(a)}(A)$ is a direct sum of the `triangular' UIR's
$(p,0)$ of $SU(3)$, for $p=0,1,2,\ldots$; and similarly 
${\cal U}^{(b)}(A)$  is a direct sum of the conjugate 
`triangular' UIR's $(0,q)$.  We indicate this by
     \begin{eqnarray}
     {\cal U}^{(a)} &=& \sum\limits^{\infty}_{p=0,1,\ldots}\oplus
     \;(p,0),\nonumber\\
     {\cal U}^{(b)} &=& \sum\limits^{\infty}_{q=0,1,\ldots}\oplus\;
     (0,q) .
     \label{3}
     \end{eqnarray}

\noindent
The total generators $Q_{\alpha}$ defined in eqn$(\ref{2})$ then
generate the product UR ${\cal U}(A)={\cal U}^{(a)}(A) \times
{\cal U}^{(b)}(A)$ on ${\cal H}$, and this is the Schwinger
UR of $SU(3)$.  It does contain every UIR $(p,q)$ of
$SU(3)$, but each one occurs an infinite number of times.
This can be seen from the Clebsch-Gordan decomposition of
the direct product $(p,0)\times (0,q)$ of two triangular UIR's\cite{9}:
     \begin{eqnarray}
     (p,0)\times (0,q) = \sum\limits^{r}_{\rho=0,1,\ldots}
     \oplus\; (p-\rho, q-\rho),\;r=\mbox{min}\;(p,q) ,
     \label{4}
     \end{eqnarray}

\noindent
which is multiplicity-free.  Applying this to each pair in the
product ${\cal U}^{(a)}\times{\cal U}^{(b)}$ we easily reach the
stated conclusion.

An efficient way to handle this infinite multiplicity is
based on the use of the semi-simple non compact Lie group
$Sp(2,R)$, more specifically some of its UIR's belonging to 
the positive discrete class\cite{10}.  In the present context the 
hermitian $Sp(2,R)$ generators and their commutation relations
are:
     \begin{mathletters}
     \begin{eqnarray}
     J_0&=& \frac{1}{2}\left(\hat{a}^{\dag}_j \hat{a}_j + 
     \hat{b}_j^{\dag} \hat{b}_j + 3\right),\nonumber\\
     K_1&=&\frac{1}{2}\left(\hat{a}_j^{\dag}\hat{b}_j^{\dag} +
     \hat{a}_j\hat{b}_j\right),\nonumber\\
     K_2 &=&\frac{-i}{2}\left(\hat{a}_j^{\dag}
     \hat{b}_j^{\dag} - \hat{a}_j \hat{b}_j\right) ;\\
     \protect[J_0,K_1\protect]&=&i\;K_2,\;
     \protect[J_0, K_2\protect]=-i\;K_1,\;
     \protect[K_1, K_2\protect] = -i\;J_0 .
     \end{eqnarray}
     \label{5}
     \end{mathletters}

\noindent
The crucial property is that the $SU(3)$ and the $Sp(2,R)$
generators mutually commute:
    \begin{eqnarray}
    [J_0\;\mbox{or}\;K_1\;\mbox{or}\;K_2,\;Q_{\alpha}]=0 .
    \end{eqnarray}

\noindent
Thus the two UR's commute as well, and $Sp(2,R)$ is just large enough
to be able to completely lift the degeneracy or multiplicity
of $SU(3)$ UIR's.  In other words, the UIR's of the product group
$SU(3)\times  Sp(2,R)$ that occur in ${\cal H}$ do so in a 
multiplicity-free manner.  This is reflected at the Hilbert space
level in the following manner.  We first decompose the individual 
Hilbert spaces ${\cal H}^{(a)}, {\cal H}^{(b)}$ into mutually 
orthogonal subspaces reflecting the decompositions $(\ref{3})$:
     \begin{eqnarray}
     {\cal H}^{(a)} &=& \sum\limits^{\infty}_{p=0,1,\ldots}\oplus
      \;{\cal H}^{(p,0)},\nonumber\\
     {\cal H}^{(b)}&=& \sum\limits^{\infty}_{q=0,1,\ldots}\oplus
     \;{\cal H}^{(0,q)}.
     \label{7}
     \end{eqnarray}

\noindent
The subspace ${\cal H}^{(p,0)}\subset {\cal H}^{(a)}$ is
of dimension $d(p,0)=\frac{1}{2}(p+1)(p+2)$; consists of
all eigenvectors in ${\cal H}^{(a)}$ of the total $a$-
type number operator $\hat{a}_j^{\dag}\hat{a}_j$ with eigenvalue 
$p$; and carries the UIR $(p,0)$ of $SU(3)$.  Similarly the
subspace ${\cal H}^{(0,q)}\subset {\cal H}^{(b)}$ is of 
dimension $d(0,q)=\frac{1}{2}(q+1)(q+2)$; consists of all
eigenvectors in ${\cal H}^{(b)}$ of the total $b$-type
number operator $\hat{b}_j^{\dag} \hat{b}_j$ with eigenvalue
$q$; and carries the UIR $(0,q)$ of $SU(3)$.  After forming the
direct product ${\cal H}^{(a)}\times {\cal H}^{(b)}$, using
eqn.$(\ref{7})$ and the Clebsch-Gordan decomposition $(\ref{4})$, we arrive 
at an orthogonal subspace decomposition for ${\cal H}={\cal H}^
{(a)}\times {\cal H}^{(b)}$:
     \begin{eqnarray}
     {\cal H}&=&\sum\limits^{\infty}_{p,q=0,1,\ldots}\oplus\;
     {\cal H}^{(p,0)}\times{\cal H}^{(0,q)}\nonumber\\
     &=&\sum\limits^{\infty}_{p,q=0,1,\ldots}
     \sum\limits^{\infty}_{\rho=0,1,\ldots}\oplus\;
     {\cal H}^{(p,q;\rho)},\nonumber\\
     {\cal H}^{(p,q;\rho)}&\subset & {\cal H}^{(p+\rho,0)}
     \times {\cal H}^{(0,q+\rho)} .
     \end{eqnarray}

\noindent
For each $\rho , {\cal H}^{(p,q;\rho)}$ is of dimension
$d(p,q)=\frac{1}{2}(p+1)(q+1)(p+q+2)$ and carries the
$\rho$th occurrence of the UIR $(p,q)$ of $SU(3)$.
For $\rho^{\prime}\neq \rho, {\cal H}^{(p,q;\rho^{\prime})}$
and ${\cal H}^{(p,q;\rho)}$ are mutually orthogonal subspaces; 
and if $p^{\prime}\neq p$ and/or $q^{\prime}\neq q$, again
${\cal H}^{(p^{\prime},q^{\prime};\rho)}$ and 
${\cal H}^{(p,q;\rho)}$ are mutually orthogonal.  An 
orthonormal basis for ${\cal H}$ consists of vectors labelled
as follows:
     \begin{eqnarray}
     |p,q &;& I, M,Y; m> :\nonumber\\
     p,q &=& 0,1,2,\ldots,\nonumber\\
     m&=& k,k+1, k+2,\ldots ,\nonumber\\
     k&=& \frac{1}{2}(p+q+3) =\frac{3}{2},2,\frac{5}{2},\ldots .
     \label{9}
     \end{eqnarray}

\noindent
Here $I,M,Y$ are `magnetic quantum numbers' within the UIR 
$(p,q)$ of $SU(3)$, with well-known ranges\cite{11}; and $m$ is the
eigenvalue of the $Sp(2,R)$ generator $J_0$.  The total numbers
of $a$-type quanta and of $b$-type quanta in the 
state displayed in eqn.$(\ref{9})$ are:
     \begin{eqnarray}
     N_a = \mbox{eigenvalue of}\;\hat{a}_j^{\dag}\hat{a}_j
     &=& p+m-k ,\nonumber\\
     N_b = \mbox{eigenvalue of}\;\hat{b}_j^{\dag}\hat{b}_j
     &=& q+m-k .
     \end{eqnarray}

\noindent
For fixed $p,q$ and $m$, as $I,M,Y$ vary within the UIR 
$(p,q)$ of $SU(3)$, we obtain an orthonormal basis for
${\cal H}^{(p,q; m-k)}$.  Switching to $\rho=m-k$ we can say:
     \begin{equation}
     {\cal H}^{(p,q;\rho)} = Sp\{|p,q;I,M,Y; k+\rho>|p,q,\rho\;
     \mbox{fixed}\;, I,M,Y\;\mbox{varying}\}
     \label{11}
     \end{equation}

\noindent
On the other hand, if we keep $p,q,I,M,Y$ fixed and let $m$ vary, 
we get an orthonormal basis for a subspace of ${\cal H}$
carrying the infinite dimensional positive discrete class UIR
$D^{(+)}_k$ of $Sp(2,R)$\cite{10}.  In other words, each of these UIR's
$D^{(+)}_k$ of $Sp(2,R)$ occurs $d(2k-3,0)+d(2k-4,1)+\ldots +
d(1,2k-4)+d(2k-3)$ times, being the sum of the dimensions of
the $SU(3)$ UIR's $(2k-3,0),(2k-4,1),\ldots (1,2k-4),(0,2k-3)$.
(The range of $2k$ is $3,4,5,\ldots$).  Since our main interest
is in UR's and UIR's of $SU(3)$, and we wish to use UIR's of 
$Sp(2,R)$ mainly to keep track of the multiplicities of the 
former, we do not introduce special notations for the
subspaces of ${\cal H}$ carrying the various $Sp(2,R)$ UIR's.
However we do note that, as stated earlier, each of the UIR's
$(p,q)\times D^{(+)}_{\frac{1}{2}(p+q+3)}$ of 
$SU(3)\times Sp(2,R)$ appears just once in ${\cal H}$, for 
$p,q=0,1,2,\ldots$.

At the generator level we can say that when the $SU(3)$ generators
$Q_{\alpha}$ act on $|p,q; I,M,Y; m>$, they alter only the  quantum
numbers $I,M,Y$ in a manner known from the representation
theory of $SU(3)$ \cite{12}; while the actions by the $Sp(2,R)$ generators
$J_0,K_1,K_2$ lead only to changes in the quantum number $m$
according to the UIR $D^{(+)}_{k}$\cite{10}.

It is in this manner that the $Sp(2,R)$ structure helps us
handle the multiplicity problem of UIR's of $SU(3)$ which is
an unavoidable feature of the Schwinger construction.  One
can now look for a natural subspace of ${\cal H},{\cal H}_0$ 
say, such that it carries every UIR $(p,q)$ of $SU(3)$ 
exactly once.  This can be done if we restrict ourselves to the
`ground state' within each $Sp(2,R)$ UIR $D^{(+)}_k$, namely if
we set $m=k$.  This amounts to  picking up the `first' occurrence
of each UIR $(p,q)$ of $SU(3)$ corresponding to $\rho=0$, or to the
`leading piece' in the reduction of each tensor product 
${\cal H}^{(p,0)}\times {\cal H}^{(0,q)}$:
     \begin{eqnarray}
     {\cal H}_0&=&\sum\limits^{\infty}_{p,q=0,1,\ldots}\oplus\;
     {\cal H}^{(p,q;0)}\nonumber\\
     &=& Sp\{|p,q;I,M,Y; k>| p,q,I,M,Y\;\mbox{varying}\}
     \nonumber\\
     &=&\{|\psi>\in\;{\cal H} | (K_1 - i\;K_2)|\psi > = 0\}.
     \label{12}
     \end{eqnarray}

\noindent
The UR of $SU(3)$ carried by ${\cal H}_0, {\cal D}_0$ say, may 
be called a Generating Representation for this group, in the
sense that each UIR is present, and exactly once:
     \begin{eqnarray}
     {\cal D}_0 =\sum\limits^{\infty}_{p,q=0,1,\ldots}\oplus\;
     (p,q) .
      \end{eqnarray}

It now turns out that just this property is also present in
the UR ${\cal D}_{SU(2)}^{({\rm ind},0)}$ of $SU(3)$ induced from
the trivial one-dimensional UIR of the canonical $SU(2)$
subgroup\cite{13}.  The corresponding Hilbert space is denoted by
${\cal H}_{SU(2)}^{({\rm ind},0)}$. (Hereafter, for simplicity, 
the superscript zero and the subscript $SU(2)$ will be omitted.) 
We can set up a one-to-one mapping 
between ${\cal H}_0$ and ${\cal H}^{({\rm ind})}$ 
preserving scalar products and $SU(3)$ actions, thus realising 
the equivalence of ${\cal D}_0$ and ${\cal D}^{({\rm ind})}$.  
First we describe ${\cal H}_0$ and ${\cal D}_0$ more explicitly. 
Denote by $|\underline{0},\underline{0}>$ the Fock vacuum in 
${\cal H}$  annihilated by $\hat{a}_j$ and $\hat{b}_j,j=1,2,3$.
Then a general vector in ${\cal H}_0$ is a collection of symmetric 
traceless tensors with respect to $SU(3)$, one for each UIR
$(p,q)$:
     \begin{mathletters}
     \begin{eqnarray}
     |\psi>\in\;{\cal H}_0&:&\nonumber\\
     |\psi>&=&\sum\limits^{\infty}_{p,q=0,1,\ldots}\;
     \psi^{j_{1}\ldots j_{p}}_{k_{1}\ldots k_{q}}\;
     \hat{a}^{\dag}_{j_{1}}\ldots \hat{a}^{\dag}_{j_{p}}\;
     \hat{b}^{\dag}_{k_{1}}\ldots \hat{b}^{\dag}_{k_{q}}\;
     |\underline{0},\underline{0}> ;\\
     &&\nonumber\\
     &&\psi^{j_{P(1)}\ldots j_{P(p)}}_{k_{Q(1)}\ldots k_{Q(q)}}=
     \psi^{j_{1}\ldots j_{p}}_{k_{1}\ldots k_{q}}\;,\;
     P\in S_p,\;Q\in S_q ;\\
     &&\nonumber\\
     &&\psi^{j\; j_{2}\ldots j_{p}}_{j \;k_{2}\ldots k_{q}} = 0\\
     &&\nonumber\\
     <\psi|\psi>&=& \parallel \psi \parallel^2 = \sum\limits^{\infty}
     _{p,q=0,1,\ldots}\; p! \;q!\;\psi^{j_{1}\ldots j_{p}}
     _{k_{1}\ldots k_{q}}\;^*
     \psi^{j_{1}\ldots j_{p}}_{k_{1}\ldots k_{q}};\\
     &&\nonumber\\
     &&{\cal D}_0(A)|\psi > = |\psi^{\prime}>,\nonumber\\
      &&\nonumber\\
     &&\psi^{\prime\;j_{1}\ldots j_{p}}
      _{k_{1}\ldots k_{q}} = A^{j_{1}}\;_{\ell_{1}}\ldots
     A^{j_{p}}\;_{\ell_{p}}\;
     A^{k_{1}}\;_{m_{1}}^* \ldots A^{k_{q}}\;_{m_{q}}^*\ldots
     \psi^{\ell_{1}\ldots \ell_{p}}_{m_{1}\ldots m_{q}} .
     \end{eqnarray}
     \label{14}
     \end{mathletters}

\noindent
Here $S_p$ and $S_q$ are the permutation groups on  $p$ and on
$q$  objects respectively.  Turning to ${\cal H}^{({\rm ind})}$ 
and ${\cal D}^{({\rm ind})}$, the former consists of complex square 
integrable functions on the
coset space $SU(3)/SU(2)$, namely the unit sphere in ${\cal C}^3$\cite{14}:
     \begin{eqnarray}
     {\cal H}^{({\rm ind})} =\left\{\psi(\underline{\xi})
     \in {\cal C}, \underline{\xi}\in{\cal C}^3|
     \parallel\psi\parallel^2 =\int \prod_{j=1}^{3}
     \left(\frac{d^2\xi_j}{\pi}\right)\delta \left(\underline{\xi}
     ^{\dag}\underline{\xi}-1\right)|\psi(\underline{\xi})|^2
     \right\}.
     \label{15}
     \end{eqnarray}

\noindent
The group action is by change of argument:
     \begin{eqnarray}
     {\cal D}^{({\rm ind})}(A)\psi&=& \psi^{\prime},\nonumber\\
     \psi^{\prime}(\underline{\xi})&=&\psi(A^{-1}\underline{\xi})
     \label{16}
     \end{eqnarray}.

\noindent
Then the one-to-one mapping between ${\cal H}_0$ and
${\cal H}^{({\rm ind})}$ consistent with the two norm
definitions $(\ref{14}d, \ref{15})$ and the two group actions 
$(\ref{14}e, \ref{16})$ is:
     \begin{eqnarray}
     |\psi> &=& \left\{\psi^{j_{1}\ldots j_{p}}_{k_{1}
     \ldots k_{q}}\right\} \in\;{\cal H}_0\longleftrightarrow
     \nonumber\\
     &&\nonumber\\
     \psi(\underline{\xi})&=& \sum\limits^{\infty}
     _{p,q=0,1,\ldots}\sqrt{(p+q+2)!}\;\;
     \psi^{j_{1}\ldots j_{p}}_{k_{1}\ldots k_{q}}\;
     \xi_{j_{1}}\ldots \xi_{j_{p}}\;
     \xi^*_{k_{1}}\ldots \xi^*_{k_{q}}\in\;
     {\cal H}^{({\rm ind})}
     \label{17}
     \end{eqnarray}

\noindent
The fact that $\psi(\underline{\xi})\in\;{\cal H}^{({\rm ind})}$
is expressible in this way in terms of \underline{traceless}
symmetric tensors is a consequence of the constraint 
$\underline{\xi}^{\dag}\underline{\xi}=1$.

In this way we see how the Schwinger UR ${\cal U}(A)$ of
$SU(3)$ contains within it a multiplicity-free UR
${\cal D}_0$ including every UIR of $SU(3)$, which is also
accessible by the method of induced representations. We will see later that 
in fact there is a continuously infinite family of subspaces 
${\cal H}_{\kappa}\subset {\cal H}$, labelled by a complex number $\kappa$, 
such that each ${\cal H}_{\kappa}$ is $SU(3)$ invariant and carries a UR 
${\cal D}_{\kappa}$ of $SU(3)$ which, like ${\cal D}_0$, is multiplicity 
free and contains each UIR $(p,q)$ without exception.

\section{Interplay between Heisenberg-Weyl and unitary group
coherent states - one and two degrees of freedom}

We now turn to an examination of the interconnections between
H-W coherent states and unitary group coherent states.  In each 
case there are both standard and generalised coherent state systems.  
In this Section we look at the cases of $n=1$ and $n=2$ degrees of
freedom, the relevant unitary groups being $U(1)$ and $SU(2)$ and
there being no multiplicity problems.  We review briefly some
known material but highlighting some special aspects.  This
material is then used as guidance when we take up in the next
Section the case $n=6$ and the Schwinger $SU(3)$ construction.

\noindent
\underline{One degree of freedom}

It is convenient to be able to switch between the use of 
non hermitian creation and annihilation operators $\hat{a}^{\dag},
\hat{a}$ and their hermitian position and momentum components 
$\hat{q},\hat{p}$:
     \begin{eqnarray}
     \hat{a}=\frac{1}{\sqrt{2}} (\hat{q} +i\hat{p}),\;
     \hat{a}^{\dag}= \frac{1}{\sqrt{2}} (\hat{q}-i\hat{p}).
     \end{eqnarray}

\noindent
For one degree of freedom, the canonical commutation relation
     \begin{eqnarray}
     \protect[\hat{a}, \hat{a}^{\dag}\protect]&=& 1,\nonumber\\
     \protect[\hat{q}, \hat{p}\protect]&=& i ,
     \label{19}
     \end{eqnarray}

\noindent
is preserved under the linear inhomogeneous transformation
     \begin{eqnarray}
     {\hat{q}\choose\hat{p}}\rightarrow {\hat{q}^{\prime}\choose
     \hat{p}^{\prime}} &=& S{\hat{q}\choose\hat{p}} +
     {q_0\choose p_0} ;\nonumber\\
     S= \left(\matrix{a&b\cr c&d}\right)
     &,& ad - bc = 1;\;q_0,p_0\in\;{\cal R}.
     \end{eqnarray}

\noindent
Here $S$ is an element of $Sp(2,R)=SL(2,R)$, and these
transformations constitute the semi direct product of $Sp(2,R)$
with the two-dimensional Abelian group of phase-space
translations.  However, as is well known, these transformations
are realised on the Hilbert space ${\cal H}$, on which
$\hat{a}^{\dag},\hat{a}$ or $\hat{q},\hat{p}$ act irreducibly, 
by unitary transformations forming a faithful UIR of a group
$G^{(1)}$ which is the semi-direct product of the
metaplectic group $Mp(2)$ with the H-W group\cite{15}:
     \begin{eqnarray}
     G^{(1)} = Mp(2) \times \{\mbox{H-W group}\}.
     \end{eqnarray}

\noindent
Each factor here is a three parameter Lie group, so $G^{(1)}$
is a six-parameter Lie group.  The H-W group is the invariant 
subgroup; it is non Abelian because of the nonzero right hand sides 
in the commutators $(\ref{19})$.  Its generators are $\hat{q},\hat{p}$
and the unit operator on ${\cal H}$.  The homogeneous part 
$Mp(2)$ is a double cover of $Sp(2,R)$; its generators are 
hermitian quadratic expressions in $\hat{a}^{\dag}$ and
$\hat{a}$, or in $\hat{q}$ and $\hat{p}$\cite{16}.  In particular the
$U(1)$ generator is $\frac{1}{2}\left(\hat{a}^{\dag}\hat{a} +\frac
{1}{2}\right)$, and this is the analogue of $J_0$ in the 
$Sp(2,R)$ Lie algebra $(\ref{5})$.

As stated above, ${\cal H}$ carries a particular UIR of $G^{(1)}$.
Upon restriction to the H-W subgroup, this representation remains 
irreducible; it is the result of exponentiating  the well-known
unique Stone-von Neumann representation of the commutation 
relations $(\ref{19})$\cite{17}.  On the other hand, upon restriction to the 
$Mp(2)$ subgroup, we get a direct sum of two UIR's of the positive 
discrete class, namely $D^{(+)}_{1/4}$ and $D^{(+)}_{3/4}$\cite{18}.
These act on the subspaces ${\cal H}^{(\pm)}$ of ${\cal H}$
consisting of even/odd parity states or Schrodinger wave functions.  
The nontrivial H-W generators $\hat{q}$ and $\hat{p}$ intertwine 
these two UIR's of $Mp(2)$.

With this background, we collect some remarks regarding various
systems of coherent states.  As both $G^{(1)}$ and the H-W
group are represented irreducibly on ${\cal H}$, for any choice
of a (normalised) fiducial vector $\psi_0\in\;{\cal H}$ we can
build up a family of $G^{(1)}$ - GCS or a family of H-W GCS\cite{5}.  These
are the orbits of $\psi_0$ under $G^{(1)}$ action and under H-W 
action respectively, and the latter orbit is a subset of the former.
In the case of $Mp(2)$, we can construct systems of GCS 
separately in ${\cal H}^{(+)}$ and in ${\cal H}^{(-)}$,
associated with any choices of fiducial vectors in these 
subspaces.  Examples are the single mode squeezed coherent states
and their variations\cite{18}.

Now let us limit ourselves to H-W coherent states, and to their
behaviours under the maximal compact $U(1)$ subgroup of $Mp(2)$.
As mentioned earlier the generator of this $U(1)$ is $\frac{1}{2}
\left(\hat{a}^{\dag}\hat{a}+\frac{1}{2}\right)$.  However for
simplicity we shall work with
     \begin{eqnarray}
     \overline{U}(\alpha) = e^{-i\alpha \hat{a}^{\dag}\hat{a}}\;,\;
     0\leq \alpha < 2\pi .
     \end{eqnarray}

\noindent
Conjugation by $\overline{U}(\alpha)$ has these effects on 
$\hat{a},\hat{a}^{\dag}$, and the unitary phase space displacement
operators $D(z)$ which represent elements of the H-W group:
     \begin{eqnarray}
     \overline{U}(\alpha)\;\hat{a}\;\overline{U}(\alpha)^{-1}&=&
     e^{i\alpha}\hat{a},\nonumber\\
     \overline{U}(\alpha)\;\hat{a}^{\dag}\;\overline{U}
     (\alpha)^{-1} &=& e^{-i\alpha}\hat{a}^{\dag} ;\nonumber\\
     D(z) &=&\exp\left(z\hat{a}^{\dag} - z^* \hat{a}\right),\nonumber\\
     \overline{U}(\alpha)\;D(z)\;\overline{U}(\alpha)^{-1} &=&
     D\left(e^{-i\alpha} z\right) .
     \label{23}
     \end{eqnarray}

\noindent
The H-W SCS correspond to the choice of the Fock vacuum
$|0>$ as the fiducial vector\cite{3}:
     \begin{equation}
     |z> = D(z)|0>\;,\;z \in\;{\cal C}.
     \end{equation}

\noindent
Invariance of $|0>$ under $\overline{U}(\alpha)$ action then 
leads to the behaviour 
     \begin{equation}
     \overline{U}(\alpha) \;|z> = |\;e^{-i\alpha} z> .
     \label{25}
     \end{equation}

\noindent
These states enjoy the well-known Klauder formula for
resolution of the identity operator:
     \begin{equation}
     \int\limits_{{\cal C}}\;\frac{d^2z}{\pi}\;|z>< z|
     = 1\;\mbox{on}\;{\cal H}.
     \label{26}
     \end{equation}

\noindent
This can be viewed as a consequence of the Schur lemma and
the square integrability of the Stone-von Neumann UIR
of the H-W group\cite{19}, since the uniform integration measure on
the complex plane in $(\ref{26})$ is essentially the invariant 
measure on the H-W group.

We now examine two variations of these familiar results.
By eqn.$(\ref{25})$, the left hand side of eqn.$(\ref{26})$ is explicitly
$U(1)$-invariant.  We can consider including some nontrivial
function $f(z^*z)$ inside the integral, which would maintain
$U(1)$ invariance, and define the operator
     \begin{equation}
     A(f) = \int\limits_{{\cal C}}\;\frac{d^2z}{\pi}
     \;f(z^*z)\;|z><z|.
     \label{27}
     \end{equation}

\noindent
As long as $f(z^*z)$ is not a constant, the integration
measure here is no longer the invariant measure on the
H-W group, so the Schur lemma is not available.  Formally,
     \begin{equation}
     f(z^*z)\neq \;\mbox{constant}\;\Longleftrightarrow
     D(z)\;A(f)\neq A(f)\;D(z) ,
     \end{equation}

\noindent
so there is no reason to expect $A(f)$ to be a multiple of the
identity.  However, $U(1)$ invariance,
     \begin{equation}
     \overline{U}(\alpha)\;A(f) = A(f)\;\overline{U}(\alpha) ,
     \end{equation}

\noindent
implies that $A(f)$ is a linear combination of projections on
to the various Fock states, and indeed we find:
     \begin{equation}
     A(f) =\sum\limits^{\infty}_{n=0}\;\int\limits^{\infty}
     _{0}\;dx \;f(x)\;x^n\;e^{-x}\cdot
     \frac{|n><n|}{n!} .
     \label{30}
      \end{equation}

\noindent
Clearly the only choice of $f$ leading to the Klauder formula $(\ref{26})$ 
is $f=1$.  On the other hand, if we choose $f(z^*z)=
\delta\left(z^*z-r^2_0\right)$ for some real positive $r_0$,
we are limiting ourselves to a subset of H-W SCS lying on a circle 
in the complex plane.  This is essentially the $U(1)$ group 
manifold; and if $r_0=1$ we have exactly the manifold
$S^1$, that is, we have a $U(1)$-worth of H-W SCS.  In this
case we find:
     \begin{eqnarray}
     f(x) &=& \delta\left(x-r^2_0\right):\nonumber\\
     A(f)&=& \int\;\frac{d^2 z}{\pi}\;\delta\left(
     z^*z-r^2_0\right)\;|z><z|\nonumber\\
     &=&\int\limits^{2\pi}_{0}\;\frac{d\theta}{2\pi}\;
     |r_0\;e^{i\theta}>< r_0\;e^{i\theta} |\nonumber\\
     &=&\sum\limits^{\infty}_{n=0}\;e^{-r^2_0}\;\frac{r_{0}^{2n}}
     {n!}\;|n><n|\nonumber\\
     &=&e^{-r^2_0}\cdot r_0^{2\hat{N}}\big/ \hat{N}! ,\nonumber\\
     \hat{N}&=& \hat{a}^{\dag}\hat{a} .
     \label{31} 
     \end{eqnarray}

\noindent
This means that even though the subset of H-W SCS 
$\left\{|r_0 e^{i\theta}>,0\leq \theta <2\pi\right\}$ lying
on a circle in the complex plane is `total'\cite{20}, and each Fock
state $|n>$ can be projected out of this subset as
     \begin{eqnarray}
     |n> = e^{r^2_0/2}\cdot \sqrt{n!}\;r_0^{-n}\cdot
     \int\limits^{2\pi}_{0}\;\frac{d\theta}{2\pi}\cdot
     e^{-in\theta}\cdot |r_0e^{i\theta}> ,
     \end{eqnarray}

\noindent
we cannot obtain a Klauder-type resolution of the identity
using them.  Thus this $U(1)$-worth of SCS does not form a 
system of GCS in the Klauder sense.

The next variation we consider is replacing the Fock vacuum
$|0>$ by a generic unit vector $|\psi_0>\in\;{\cal H}$ as fiducial 
vector.  We then get a family of H-W GCS\cite{21}:
     \begin{equation}
     |z;\psi_0> = D(z)\;|\psi_0>,\;z\in\;{\cal C}.
     \end{equation}

\noindent
Once again, Schur lemma leads to the Klauder resolution
of the identity,
     \begin{equation}
     \int\;\frac{d^2z}{\pi}\;|z;\psi_0>< z;\;\psi_0| = c.1,
     \end{equation}

\noindent
for some constant $c$; and square integrability ensures that $c$ is 
finite. If in the manner of eqn.$(\ref{27})$ we next define
     \begin{equation}
     A(f;\psi_0) = \int\;\frac{d^2z}{\pi}\;f(z^*z)
      \;|z;\psi_0><z;\psi_0|,
     \end{equation}

\noindent
then on the one hand we do not expect $A(f;\psi_0)$ to be a multiple of
the unit operator since we lose Schur lemma; and on the other
hand we do not even expect $A(f;\psi_0)$ to  commute with
$\overline{U}(\alpha)$.  That is, in general $A(f;\psi_0)$ is
not a linear combination of the projections $|n><n|$ on to the
Fock states.  The exceptions are when $|\psi_0>$ is an eigenstate 
of $\hat{a}^{\dag}\hat{a}$, ie., a Fock state $|n_0>$ for some
integer $n_0$.  This possibility arises because $U(1)$ is
Abelian, and its UIR's are all one-dimensional.  In that case we
find\cite{22}:
     \begin{eqnarray}
     |\psi_0> &=& |n_0>:\nonumber\\
     \overline{U}(\alpha) |z;n_0> &=& e^{-i\alpha n_0}|
     e^{-i\alpha} z; n_0>,\nonumber\\
     \overline{U}(\alpha) A(f;n_0) &=& A(f;n_0) \overline{U}
     (\alpha);\nonumber\\
     A(f;n_0)&=& \sum\limits^{\infty}_{n=0}\;C_{n,n_0}(f)\;
     |n><n|,\nonumber\\
     C_{n,n_0}(f) &=& \frac{n_<!}{n_>!}\;\int\limits^{\infty}
     _0\;dx \;f(x)\;x^{|n-n_0|}
     e^{-x}\left(L_{n_{<}}^{|n-n_0|}(x)\right)^2 ,\nonumber\\
     n_>&=& \mbox{max}\;(n,n_0), n_< =\mbox{min}\;(n,n_0) .
     \label{36}
     \end{eqnarray}

\noindent
When $n_0=0$ we recover eqn.$(\ref{30})$.  If we next choose 
$f(z^*z)=\delta\left(z^*z-r^2_0\right)$, thus
limiting ourselves to a $U(1)$-worth of H-W GCS, we find 
in place of eqn.$(\ref{31})$):
     \begin{eqnarray}
     f(x)&=& \delta\left(x-r^2_0\right):\nonumber\\
     A(f;n_0)&=& \int\;\frac{d^2z}{\pi}\;\delta
     \left(z^*z-r^2_0\right) |z;n_0><z;n_0|\nonumber\\
     &=&\int\limits^{2\pi}_{0}\;\frac{d\theta}{2\pi}\;
     |r_0 e^{i\theta};n_0>< r_0 e^{i\theta};n_0|\nonumber\\
     &=&\sum\limits^{\infty}_{n=0}\;e^{-r^2_0} r_0^{2n}
     \left(L_{n_{<}}^{|n-n_0|} \left(r^2_0\right)\right)^2\;
     \frac{n_<!}{n_>!}\;|n><n| .
     \end{eqnarray}

The main result of these considerations is that with
SCS or GCS for the H-W group for one degree of freedom, we
can get a Klauder type resolution of the identity only if we
use the invariant measure on the group, but understandably 
not if we limit ourselves to a subset amounting to a 
$U(1)$-worth of these states.

\noindent
\underline{Two degrees of freedom}

Here we are interested in the interplay between coherent state 
systems for the relevant five-parameter H-W group, and the 
unitary groups $U(2)$ and $SU(2)$ which were the subject of 
the original Schwinger construction.  

The non vanishing commutators in non hermitian and hermitian
forms are
     \begin{eqnarray}
     \protect[\hat{a}_r, \hat{a}_s^{\dag}\protect]&=&      
     \delta_{rs},\nonumber\\
     \protect[\hat{q}_r, \hat{p}_s\protect]
     &=& i\;\delta_{rs} ,\;\; r,s=1,2.
    \label{38}
    \end{eqnarray}

\noindent
There is no cause for confusion if again we write 
${\cal H}$ for the Hilbert space carrying the irreducible 
Stone-von Neumann representation of these relations.
The largest natural invariance group now acts on the four
$\hat{q}$'s and $\hat{p}$'s as follows:
     \begin{eqnarray}
     {\hat{q}_r\choose \hat{p}_r}\longrightarrow
     {\hat{q}^{\prime}_r\choose \hat{p}^{\prime}_r}
      = S{\hat{q}_r\choose\hat{p}_r} + {q_{r,0}\choose
      p_{r,0}}.
      \end{eqnarray}

\noindent
Here $S\in\;Sp(4,R)$ is a four-dimensional real symplectic
matrix, and $q_{r,0},p_{r,0}$ denote an Abelian phase space
translation\cite{23}.  These fourteen parameter transformations
preserve $(\ref{38})$.  They make up the semi direct product of
$Sp(4,R)$, which is ten dimensional, with the four 
dimensional Abelian translations.  On the space ${\cal H}$,
however, these transformations are realised as a faithful UIR 
of the fifteen-parameter semi direct product
     \begin{eqnarray}
     G^{(2)} = Mp(4) \times \{\mbox{H-W group}\}.
     \end{eqnarray}

\noindent
Here the invariant subgroup is the five-parameter non Abelian
H-W group appropriate for two degrees of freedom, while the
homogeneous part is the metaplectic group $Mp(4)$, a double
cover of $Sp(4,R)$.  The generators of the former are $\hat{q}_r,
\hat{p}_r$ and the unit operator, while those of the latter are
hermitian symmetrised quadratics in $\hat{q}_r,\hat{p}_r$.

The Hilbert space ${\cal H}$ carries a UIR of $G^{(2)}$,
which remains irreducible when restricted to the H-W group.  
On the other hand, $Mp(4)$ is represented by the direct sum
of two UIR's, one each on the subspaces of even and odd
parity states in ${\cal H}$.  The general statements that
can be made about GCS with respect to $G^{(2)},Mp(4)$ and
the H-W group are similar to those in the one degree of freedom
case.  Once again, our main interest is in the connections
between H-W and $SU(2)$ coherent state systems.

The maximal compact subgroup of $Mp(4)$ is $U(2)$.  The
$SU(2)$ part of $U(2)$ has the generators and commutation 
relations ( Schwinger construction)
     \begin{eqnarray}
     J_j &=& \frac{1}{2} \hat{a}^{\dag} \sigma_j \hat{a},\nonumber\\
     \protect[J_j, J_k\protect]&=& i\;\epsilon_{jk\ell} 
      J_{\ell},\;\;j,k=1,2,3 .
     \end{eqnarray}

\noindent
The $U(1)$ part of $U(2)$ has as generator the total number
operator
     \begin{eqnarray}
     \hat{N}&=&\hat{N}_1 + \hat{N}_2,\nonumber\\
     \hat{N}_r &=& \hat{a}^{\dag}_r \hat{a}_r ,\nonumber\\
     \protect[J_j, \hat{N}\protect]&=&0.
     \end{eqnarray}

\noindent
For general $u\in U(2)$, we write $\overline{U}(u)$ for the
corresponding unitary operator on ${\cal H}$, generated by 
$J_j,\hat{N}$.  Then in place of eqn.$(\ref{23})$ we now have:
     \begin{eqnarray}
     \overline{U}(u) \hat{a} \overline{U}(u)^{-1} &=&
     u^{-1} \hat{a} ,\nonumber\\
     \overline{U}(u) \hat{a}^{\dag} \overline{U}(u)^{-1}
     &=& \hat{a}^{\dag} u;\nonumber\\
     D(\underline{z})&=&\exp \left(\hat{a}^{\dag}
     \underline{z} - \underline{z}^{\dag} \hat{a}\right),
     \nonumber\\
     \overline{U}(u) D(\underline{z}) \overline{U}(u)^{-1}
     &=& D(u \underline{z}).
     \label{43}
     \end{eqnarray}

\noindent
Here $\underline{z}=(z_1,z_2)^T$ is a complex two-component
column vector, while $\hat{a}$ and $\hat{a}^{\dag}$ are
written as column and row vectors respectively.

The reduction of $\overline{U}(u)$ into UIR's is
accomplished by the break-up of ${\cal H}$ into the mutually
orthogonal eigenspaces ${\cal H}^{(j)}$ of $\hat{N}$
with eigenvalues $2j$, where $j=0,1/2,1,\ldots$.  The
orthonormal Fock basis for ${\cal H}$ is made up of the 
simultaneous eigenvectors of $\hat{N}_1$ and $\hat{N}_2$:
     \begin{eqnarray}
     |n_1, n_2> &=& \frac{\left(\hat{a}^{\dag}_1\right)^{n_{1}}
      \left(\hat{a}_2^{\dag}\right)^{n_{2}}}{\sqrt{n_{1}!n_{2}!}}
      \;|0,0> ,\nonumber\\
     &&\nonumber\\
      \hat{N}_r \;|n_1,n_2> &=& n_r \;|n_1,n_2>,\;\;r=1,2 .
      \end{eqnarray}

\noindent
For the purposes of reduction of $\overline{U}$, with no
danger of confusion we use vectors labelled $|j,m>$ and defined
in terms of these Fock states by
     \begin{eqnarray}
     |j,m> &=& |n_1,n_2>,\nonumber\\
     n_1 &=& \frac{1}{2} (j+m), n_2 = \frac{1}{2}(j-m),\nonumber\\
     j&=&0, 1/2, 1, \ldots,\;\;m=j,j-1,\ldots, -j .
     \end{eqnarray}

\noindent
Then the subspaces ${\cal H}^{(j)}$ are given by
     \begin{eqnarray}
     {\cal H}^{(j)} &=& Sp\{|j,m>|j\;\mbox{fixed},\;m=j, j-1,\ldots,
     -j\}\nonumber\\
     && j=0, 1/2, 1, \ldots .
     \end{eqnarray}

\noindent
The operators $\overline{U}(u)$ leave each ${\cal H}^{(j)}$, of 
dimension $(2j+1)$, invariant, and  reduce thereon to the
spin $j$ UIR of $SU(2)$, along with the value $2j$ for the
$U(1)$ generator $\hat{N}$.  This is the known multiplicity-
free reduction of the $SU(2)$ Schwinger construction\cite{2}.  The
projection operator $P_j$ onto the subspace ${\cal H}^{(j)}$,
which will be needed later, is
     \begin{eqnarray}
     P_j =\sum\limits^{+j}_{m=-j}\;|j,m><j,m| = \delta_{\hat{N},2j}.
     \end{eqnarray}

The H-W SCS use the Fock vacuum $|0,0>$ as the fiducial vector:
     \begin{eqnarray}
     |\underline{z}> = D(\underline{z})\;|0,0>,
     \end{eqnarray}

\noindent
and on account of eqn.$(\ref{43})$ they have the $U(2)$ behaviour
     \begin{equation}
     \overline{U}(u) |\underline{z}>  = |\;u\;\underline{z}>.
     \label{49}
     \end{equation}

\noindent
This is because $|0,0>$ is invariant under $U(2)$ action;
in fact it is the only such vector in ${\cal H}$.  Therefore the 
general H-W SCS $|\underline{z}>$ is obtainable by suitable
$U(2)$ action from a SCS for the first degree of freedom
alone:
     \begin{eqnarray}
     |\underline{z}>&=& \overline{U}(u)\;|\underline{z}^{(0)}>,\;     
     \mbox{suitable}\; u\in\;U(2),\nonumber\\
     \underline{z}^{(0)} &=& r{1\choose 0} ,\nonumber\\
     r^2 &=& \underline{z}^{\dag}\underline{z},\;\;
     0\leq r< \infty .
     \label{50}
     \end{eqnarray}

To bring out the connection between these H-W SCS and 
$SU(2)$ SCS (identified below) in the clearest possible
manner, we parametrise $\underline{z}$ and define elements
$A(\theta,\phi)\in SU(2)$ in a coordinated manner:
     \begin{eqnarray}
     \underline{z}&=& e^{i\alpha} A(\theta,\phi) 
     \underline{z}^{(0)},\nonumber\\
     A(\theta,\phi)&=& e^{\frac{-i}{2}\phi \sigma_3}
     e^{\frac{-i}{2}\theta\sigma_2} \in\;SU(2),\nonumber\\
     && 0\leq \theta\leq \pi,\;\;0\leq \alpha,\;\phi
     \leq 2\pi;\nonumber\\
     z_1&=& r\;e^{i\alpha} e^{-i\phi/2} \cos\theta/2,\;
     z_2 = r\;e^{i\alpha} e^{i\phi/2}\sin\theta/2 .
     \label{51}
     \end{eqnarray}

\noindent
We view $\theta,\phi$ as spherical polar angles on $S^2$.  Then
eqn.$(\ref{50})$ assumes the more detailed form
     \begin{eqnarray}
      |\underline{z}>&=& e^{i\alpha\hat{N}}\overline{U} 
      (A(\theta,\phi))|\underline{z}^{(0)}>,\nonumber\\
      |\underline{z}^{(0)}> &=&
      e^{r\left(\hat{a}^{\dag}_1 - \hat{a}_1\right)} 
      |0,0>\nonumber\\
      &=& e^{-\frac{1}{2}\;r^2}\sum\limits^{\infty}
      _{j=0,1/2,1,\ldots} \frac{r^{2j}}{\sqrt{2j!}}\;
      |j,j> .
      \label{52}
      \end{eqnarray}

\noindent
The component of $|\underline{z}^{(0)}>$ within ${\cal H}^{(j)}$
is a multiple of $|j,j>$, the highest weight vector in the spin
$j$ UIR of $SU(2)$.  By definition, the $SU(2)$ SCS in any UIR
are based on the choice of highest weight vector (or any $SU(2)$
transform of it) as fiducial vector\cite{24}.  This vector is the 
eigenvector of $J_3$ with maximum eigenvalue $j$, so any
$SU(2)$ transform of it is an eigenvector of a suitable
combination of $J_k$ with the same (maximum) eigenvalue.
These remarks lead to the following notations for $SU(2)$
SCS:
     
     \[\overline{U}(A(\theta,\phi))\;|j,j> \equiv 
     |j,\hat{n}(\theta,\phi)>\]
     
     \[=\sum\limits^{j}_{m=-j}\;
     \sqrt{\frac{2j!}{(j+m)! (j-m)!}}\;e^{-im\phi}
     (\cos\theta/2)^{j+m}(\sin\theta/2)^{j-m}|j,m>,\]
     \begin{eqnarray}
     \hat{n}(\theta,\phi)\cdot {\buildrel\rightarrow \over J}\;
     |j,\hat{n}(\theta,\phi)>&=& j|j,\hat{n}(\theta,\phi)>,
     \nonumber\\
     \hat{n}(\theta,\phi)=(\sin\theta \cos\phi, \sin\theta
     \sin\phi, \cos\theta) &=& \frac{1}{r^2}\; \underline{z}^{\dag}
      \underline{\sigma}\; \underline{z}\;\in\; S^2 .
      \label{53}
      \end{eqnarray}

\noindent
Thus the family of $SU(2)$ SCS in the spin $j$ UIR is
$\{|j,\hat{n}(\theta,\phi)>\}$, one for each point on $S^2$ which
is the coset space $SU(2)/U(1)$.  For these states we have
the well-known properties
     \begin{mathletters}
     \begin{eqnarray}
     <j,\hat{n} (\theta^{\prime},\phi^{\prime})|
      j, \hat{n}(\theta,\phi)> &=& 
     \left(\cos\theta^{\prime}/2 \;\cos\theta/2\;e^{i(\phi^{\prime}
     -\phi)/2} +\sin\theta^{\prime}/2\;\sin\theta/2\;
     e^{i(\phi-\phi^{\prime})/2}\right)^{2j} ,\\
     A\in\;SU(2) : \overline{U}(A) |j,\hat{n}>&=&
     e^{i\omega(A;\hat{n})} |j,\;
     R(A)\hat{n}> ,
     \end{eqnarray}
     \label{54}
      \end{mathletters}

\noindent
where $R(A)\in\;SO(3)$ is the image of $A\in\;SU(2)$ under
the $SU(2)\rightarrow SO(3)$ homomorphism, and $\omega(A;\hat{n})$
is a (Wigner) phase angle\cite{25}.  Combining eqns.$(\ref{52},\ref{53})$  
we get the connection between H-W and $SU(2)$ SCS:
     \begin{eqnarray}
     |\underline{z}> = e^{-\frac{1}{2}\;r^2}\;
     \sum\limits^{\infty}_{j=0,1/2,1,\ldots}\;\frac
     {(r\;e^{i\alpha})^{2j}}{\sqrt{2j!}}\;
     |j,\hat{n}(\theta,\phi)>.
     \label{55}
     \end{eqnarray}

\noindent
We trace this direct connection to the simple $U(2)$ action
$(\ref{49})$, and the expansion $(\ref{52})$ of $|\underline{z}^{(0)}>$ in
terms of $SU(2)$ highest weight states.

We now look at the Klauder resolution of unity for the H-W SCS,
highlighting the $SU(2)$ SCS structure.  Using the parametrisation
$(\ref{51})$ for $\underline{z}$ we find:
     \begin{eqnarray*}
      \int \;\frac{d^2z_1}{\pi}\;\frac{d^2z_2}{\pi}\;
      |\underline{z}><\underline{z}| =
      \frac{1}{4\pi^2}\;\int\limits^{\infty}_{0}\;r^3 dr\;
      \int\limits^{2\pi}_{0}\;d\alpha\;
      \int\limits_{S^2}\;d\Omega (\theta,\phi)|\;
      \underline{z}><\underline{z}|
      \end{eqnarray*}
      \begin{eqnarray*}
      =\frac{1}{4\pi^2}\;\int\;r^3 dr\;d\alpha\;d\Omega
       (\theta,\phi)\;\sum\limits^{\infty}_{j,j^{\prime}=0,
       1/2,1,\ldots}&& e^{-r^2}  r^{2(j+j^{\prime})}
       e^{2i\alpha(j-j^{\prime})}\;\times\nonumber\\
       &&|j,\hat{n}(\theta,\phi)><j^{\prime},\hat{n}
       (\theta,\phi)|/\sqrt{2j!\;2j^{\prime}!}
       \end{eqnarray*}
       \begin{eqnarray}
       =\frac{1}{2\pi}\;\sum\limits^{\infty}_{j=0,1/2,1,\ldots}
       \;\frac{1}{2j!}\;\int\limits^{\infty}_{0}\;r^3 dr\;
       e^{-r^{2}} r^{4j}\;\int\limits_{S^2}\;d\Omega(\theta,\phi)
       \;|j,\hat{n}(\theta,\phi)><j,\hat{n}(\theta,\phi)|.
       \label{56}
       \end{eqnarray}

\noindent
Here $d\Omega(\theta,\phi)$ is the element of solid angle on
$S^2$.  Using eqn.$(\ref{54}b)$ we see that the integral over $S^2$ 
results in an operator invariant under the spin $j$  UIR of
$SU(2)$ appearing on ${\cal H}^{(j)}$, therefore by Schur 
lemma for this UIR we have: 
     \begin{eqnarray}
     \int\limits_{S^2}\;d\Omega (\theta,\phi)\;|j,\hat{n}
     (\theta,\phi)><j,\hat{n}(\theta,\phi)| =
     \frac{4\pi}{2j+1}\;P_j.
     \end{eqnarray}

\noindent
Substituting this in eqn.$(\ref{56})$  we get
     \begin{eqnarray}
     \int\;\frac{d^2z_1}{\pi}\;\frac{d^2z_2}{\pi}\;|
     \underline{z}><\underline{z}|&=&
     2\;\sum\limits^{\infty}_{j=0,1/2,1,\ldots}\;
     \frac{1}{(2j+1)!}\;\int\limits^{\infty}_0
     \;r^3 dr\;e^{-r^{2}}\cdot r^{4j}\cdot P_j\nonumber\\
     &=&\sum\limits^{\infty}_{j=0,1/2,1,\ldots}\;P_j
     \nonumber\\
     &=&1\;\mbox{on}\;{\cal H}.
     \end{eqnarray}

\noindent
This is known and expected on account of the Schur lemma for the
H-W UIR, since the integration measure is the invariant one
on the H-W group.  At the same time we can immediately trace 
the consequences of modifying the measure in a $U(2)$-invariant way, 
when we lose the possibility of using the lemma for the H-W UIR:
     \begin{eqnarray}
     A(f) &=& \int\;\frac{d^2z_1}{\pi}
     \;\frac{d^2z_2}{\pi}\;f(\underline{z}^{\dag}
     \underline{z})\;|\underline{z}><\underline{z}|\nonumber\\
     &&\nonumber\\  
     &=&\sum\limits^{\infty}_{j=0,1/2,1,\ldots}\;\int\limits
     ^{\infty}_0\;dx \;f(x) x^{2j+1}\;e^{-x}\;
     \frac{P_j}{(2j+1)!} ,\nonumber\\
     &&\nonumber\\
     f&\neq & \mbox{constant}\Longleftrightarrow D(\underline{z})
     \;A(f)\neq A(f)\;D(\underline{z}).
     \label{59}
     \end{eqnarray}

\noindent
With the particular choice $f(x) =\delta\left(x-r^2_0\right)$ for
real positive $r_0$, we limit ourselves to an ``$SU(2)$-worth''
of H-W SCS, and in that case we have:
     \begin{eqnarray}
     f(x) &=& \delta\left(x-r^2_0\right):\nonumber\\
      &&\nonumber\\
     A(f)&=& \int\;\frac{d^2z_1}{\pi}\;\frac{d^2z_2}{\pi}
     \;\delta\left(\underline{z}^{\dag}\underline{z} -
     r^2_0\right)\;|\underline{z}><\underline{z}|\nonumber\\
     \nonumber\\
     &=&r^2_0\;\int\limits^{2\pi}_{0}\;\frac{d\alpha}{2\pi}\cdot
     \int\limits_{S^2}\;\frac{d\Omega(\theta,\phi)}{4\pi}\cdot
     \bigg|e^{i\alpha}A(\theta,\phi){r_0\choose 0}\rangle\langle
     e^{i\alpha} A(\theta,\phi){r_0\choose 0}\bigg|\nonumber\\
     \nonumber\\
     &=&\sum\limits^{\infty}_{j=0,1/2,1,\ldots}\;
     e^{-r^2_0}\cdot \;\frac{\left(r^2_0\right)^{2j+1}}
     {(2j+1)!}\;P_j .
     \label{60}
     \end{eqnarray}

\noindent
The structure of these results $(\ref{59},\ref{60})$ is as expected since
$A(f)$ does commute with $\overline{U}(u)$.

Lastly we consider briefly some aspects of H-W GCS in the 
case of two degrees of freedom.  These arise by replacing the
Fock vacuum $|0,0>$ by some other (normalised) vector
$|\psi_0>\in\;{\cal H}$ as fiducial vector:
     \begin{eqnarray}
     |\underline{z};\psi_0> = D(\underline{z})\;|\psi_0>.
     \end{eqnarray}

\noindent
Schur lemma and square integrability of the H-W UIR ensure
the Klauder formula
     \begin{eqnarray}
     \int\;\frac{d^2z_1}{\pi}\;\frac{d^2z_2}{\pi}\;|\underline{z};
     \psi_0><\underline{z}; \psi_0| = c.1,
     \label{62}
     \end{eqnarray}
  
\noindent
for some finite constant $c$.  However, if $|\psi_0>\neq |0,0>$, 
we never have any simple behaviour for these GCS under $U(2)$ action.  
This is in contrast to eqn.$(\ref{36})$ in the case of one degree of 
freedom.  The reason is that the only one-dimensional UIR of $SU(2)$ is the 
trivial UIR, all others are of dimension two or greater.  This can be
traced to the non Abelian nature of $SU(2)$, in contrast to 
$U(1)$.  For this reason we are unable to obtain $|\underline{z};
\psi_0>$ for general $\underline{z}$ from some specially
chosen and simpler state $|\underline{z}
^{(0)}; \psi_0>$ via $U(2)$ action; so the possibility of
relating H-W GCS to some sequence of $SU(2)$ GCS's within each 
subspace ${\cal H}^{(j)}$ is also lost. Going one step further,
if we consider a modified $U(2)$-invariant measure in place
of the translation invariant one in eqn.$(\ref{62})$, but for a GCS
system, and if we define
     \begin{eqnarray}
     A(f;\psi_0) = \int\;\frac{d^2z_1}{\pi}\;\frac{d^2z_2}{\pi}
     \;f(\underline{z}^{\dag}\underline{z})\;
     |\underline{z}; \psi_0><\underline{z};\psi_0|,
     \end{eqnarray}

\noindent
for $|\psi_0> \neq |0,0>$, this will 
not commute with $\overline{U}(u)$ and will not reduce to a linear 
combination of the projections $P_j$.

\section{Relation between H-W and $SU(3)$ SCS, 
restriction to ${\cal H}_0$}

Now that we have explored the relationships between H-W SCS and
unitary group SCS for one and two degrees of freedom, we 
proceed to the $SU(3)$ Schwinger construction recalled in
Section 2, and the corresponding H-W SCS for six oscillators.  
Here we invert the order of development as compared to the
previous Section.  We recall first the definition of $SU(3)$
SCS within each UIR, then proceed to the H-W system.  The
specific new feature is the multiplicity problem, to be
handled using $Sp(2,R)$.

\noindent
\underline{$SU(3)$ Standard Coherent States}

The familiar orthonormal basis states within the UIR 
$(p,q)$ of $SU(3)$, corresponding to the canonical subgroup
chain $U(1)\subset U(2)\subset SU(3)$, consist of a set
of isospin-hypercharge multiplets (cf.eqns.$(\ref{9},\ref{11}))$\cite{26}:
     \begin{eqnarray}
     |p,q &;& I M Y\rangle ,\nonumber\\
     I&=& \frac{1}{2} (r+s),\;Y = 
     \frac{2}{3}(q-p) +r-s,\nonumber\\
     M&=& I,I-1,\ldots,\;-I+1, -I,\nonumber\\
     0&\leq &r\leq p,\;0\leq s\leq q .
     \label{64}
     \end{eqnarray}

\noindent
The highest weight state is the one with maximum possible
value of $M$:
     \begin{eqnarray}
     |p,q; \frac{1}{2}(p+q),\;\frac{1}{2}(p+q),\;
     \frac{1}{3}(p-q)\rangle .
     \label{65}
     \end{eqnarray}

\noindent
In terms of the realisation of the UIR $(p,q)$ via 
irreducible tensors $T=\left\{T^{j_{1}\ldots j_{p}}
_{k_{1}\ldots k_{q}}\right\}$, this state corresponds to the
component
      \begin{equation}
      T^{11\ldots 1}_{22\ldots 2} .
      \label{66}
      \end{equation}

\noindent
From this one can see that the stability group (upto phase 
factors) of the state $(\ref{65})$ is a subgroup $H\subset SU(3)$
dependent on $p$ and $q$.  Disregarding the trivial UIR
$(0,0)$, we have:
     \begin{mathletters}
     \begin{eqnarray}
     p\geq 1, q=0&:& H=U(2)\;\mbox{on dimensions}\; 2,3 ;\\
     p=0, q\geq 1&:&H=U(2)\;\mbox{on dimensions}\;1,3 ;\\
     p,q\geq 1 &:& H=\mbox{diagonal subgroup of}\;SU(3)
     \end{eqnarray}
     \label{67}
     \end{mathletters}

\noindent
(Here the dimensions 1,2,3 refer to the space of the
defining UIR $(1,0)$).  In eqn.$(\ref{67}a)$ (eqn.$(\ref{67}b)$), a
$U(2)$ transformation on dimensions 2 and 3 (1 and 3) is 
to be accompanied by a phase change in dimension 1(2)
to preserve unimodularity of the $SU(3)$ transformation.
The dimensionalities of these three stability groups
are four, four and two respectively.

The $SU(3)$ SCS within the UIR $(p,q)$ are the states
obtained by acting with all $SU(3)$ elements on the
highest weight state $(\ref{65})$.  They may be written as
$|p,q; A>,\;A\in\;SU(3)$:
     \begin{eqnarray}
     |p,q;A> = \overline{U}(A) |p,q; \frac{1}{2}(p+q),\;
     \frac{1}{2}(p+q), \frac{1}{3}(p-q)>.
     \label{68}
     \end{eqnarray}

\noindent
Therefore in the UIR's $(p,0)$ and $(0,q)$, they form four-
parameter continuous families of normalised states; while in
$(p,q)$ with $p,q\geq 1$ we have six-parameter continuous
families.  Referring to eqn.$(\ref{67})$ we have:
     \begin{equation}
     h\in\;H: |p,q;Ah> = e^{i\varphi(h)}
     |p,q; A> ,
     \end{equation}
 
\noindent
for some phase $\varphi(h)$.

These $SU(3)$ SCS have been studied in detail in ref.\cite{27}   ,
individually within each UIR.  As we see below, the
Schwinger construction helps us generate them collectively 
and explore some of their properties in an efficient manner, 
just as in eqn.$(\ref{55})$ we have a construction of the $SU(2)$
SCS in all its UIR's at one stroke.

If within the UIR $(p,q)$ we choose as fiducial vector
some vector other than the highest weight vector $(\ref{65})$ or
any $SU(3)$ transform of it, then we obtain a family of
$SU(3)$ GCS.  For the present we consider only SCS's,
turning to particular GCS's in subsequent Sections.

In the Hilbert space ${\cal H}$ of the $SU(3)$ Schwinger
construction the `first' occurrence of the UIR $(p,q)$
is in the subspace ${\cal H}^{(p,q;0)}\subset {\cal H}_0$
which is annihilated by $K_-$.  The corresponding highest 
weight state $(\ref{65})$, using the complete notation of eqn.$(\ref{9})$ 
and recalling eqn.$(\ref{66})$, is:
     \[|p,q; \frac{1}{2}(p+q),\;\frac{1}{2}(p+q),
     \frac{1}{3}(p-q);\;\frac{1}{2}(p+q+3)> =\]
     \begin{eqnarray}
     \frac{\left(\hat{a}^{\dag}_1\right)^p
     \left({\hat b}_2^{\dag}\right)^q}
     {\sqrt{p! q!}}\;|\underline{0},\underline{0}>\in\;
     {\cal H}^{(p,q;0)}\subset {\cal H}_0.
     \end{eqnarray}

\noindent
It follows that all these highest weight states, one for
each UIR $(p,q)$, are generated by the special H-W SCS
     \[|z_1,0,0; 0,w_2,0> =
      D(z_1,0,0,0,w_2,0)|\underline{0},\underline{0}>
      \in\;{\cal H}_0 ,\]
      \begin{eqnarray}
      D(\underline{z},\underline{w})&=&\exp
      \left(\underline{z}\cdot\underline{\hat{a}}^{\dag}
      -\underline{z}^*\cdot \underline{\hat{a}}
      +\underline{w}\cdot\underline{\hat{b}}^{\dag}
      -\underline{w}^*\cdot\underline{\hat{b}}\right)
     \nonumber\\
      &=&\exp \left(-\frac{1}{2}\underline{z}^{\dag}\underline{z}
      \;-\frac{1}{2}\underline{w}^{\dag}\underline{w}\;
      +\underline{z}\cdot\underline{\hat{a}}^{\dag}\;
      +\underline{w}\cdot\underline{\hat{b}}^{\dag}\right) .
      \end{eqnarray}

\noindent
Here $\underline{z}$ and $\underline{w}$ are independent complex 
3-vectors, and $D(\underline{z},\underline{w})$ are the 
displacement operators for the six-oscillator system of the
Schwinger construction.  Indeed we have:
     \begin{eqnarray*}
     |z_1, 0, 0; 0,w_2, 0>&=&
     e^{-\frac{1}{2}|z_1|^2 -\frac{1}{2}|w_2|^2}
     \sum\limits^{\infty}_{p,q=0}\;
     \frac{z_1^p w^q_2}{\sqrt{p! q!}}\;\times
     \end{eqnarray*}
     \begin{eqnarray}
     |p,q;\frac{1}{2}(p+q), \frac{1}{2}(p+q),
     \frac{1}{3}(p-q); \frac{1}{2}(p+q+3)>,
     \label{72}
     \end{eqnarray}

\noindent
which is analogous to the second of eqns.$(\ref{52})$.  We will use
this below.

\noindent
\underline{$SU(3)$ analysis of the H-W SCS}

For the six oscillator system used in the Schwinger $SU(3)$ 
construction the H-W SCS are labelled by two complex 
three-dimensional vectors $\underline{z}$ and $\underline{w}$,
thus the pair $(\underline{z},\underline{w})$ is a point in
${\cal C}^6$.  They are obtained by applying the 
displacement operators $D(\underline{z},\underline{w})$ to
the Fock vacuum $|\underline{0},\underline{0}>$ as fiducial
vector:
     \begin{equation}
     |\underline{z},\underline{w}> = D(\underline{z},
     \underline{w})\;|\underline{0},\underline{0}>.
     \end{equation}

\noindent
We see from eqn.$(\ref{5})$ that they are eigenstates of the
$Sp(2,R)$ lowering operator $K_-=K_1 -iK_2$:
     \begin{eqnarray}
     \hat{a}_j |\underline{z},\underline{w}>&=&
     z_j|\underline{z},\underline{w}>,\nonumber\\
     \hat{b}_j |\underline{z},\underline{w}>&=&
     w_j|\underline{z},\underline{w}>,\nonumber\\
     K_- |\underline{z},\underline{w}>&=&
     \underline{z}^T\underline{w} |\underline{z},\underline{w}>.
     \label{74}
     \end{eqnarray}

\noindent
Therefore only those SCS $|\underline{z}, \underline{w}>$ for which
$\underline{z}^T \underline{w}=0$ belong to ${\cal H}_0$.
The complete set of SCS obeys the Klauder resolution of the 
identity,
     \begin{eqnarray}
     \int\limits_{{\cal C}^{6}}\prod\limits^{3}_{j=1}
     \left(\frac{d^2z_j}{\pi}\;\frac{d^2w_j}{\pi}\right)\;
     |\underline{z},\underline{w}\rangle\langle\underline{z},
     \underline{w}| = 1\;\mbox{on}\;{\cal H} ,
     \label{75}
     \end{eqnarray}

\noindent
the integration measure being the invariant one on the
H-W group.

We  now explore the behaviour of these SCS  under
$SU(3)$ action.  From the manner in which the generators
$Q_{\alpha}$ are constructed in eqn.$(\ref{2})$ we have:
     \begin{eqnarray}
     A\;\in\;SU(3) : {\cal U}(A)\;D(\underline{z},\underline{w})\;
     {\cal U}(A)^{-1} = D(A\underline{z}, A^*\underline{w}),
     \end{eqnarray}

\noindent
from which it follows that
     \begin{eqnarray}
     {\cal U}(A) |\underline{z}, \underline{w}> =
     |A \underline{z}, A^* \underline{w}>.
     \end{eqnarray}

\noindent
The independent invariants under this action are 
$\underline{z}^{\dag} \underline{z}, \underline{w}^{\dag} 
\underline{w}$ and $\underline{z}^T \underline{w}$,
the last being the eigenvalue of $K_-$.
We describe them using four real independent parameters
$u,v,x,y$ as
     \begin{eqnarray}
     \underline{z}^{\dag} \underline{z}  = u^2&,&
     \underline{w}^{\dag} \underline{w} = v^2\;,\;
     \underline{z}^T \underline{w} = uv (x+iy),\nonumber\\
     u,v\geq 0 &,& 0\leq x^2 + y^2 \leq 1.
     \label{78}
     \end{eqnarray}

\noindent
The upper bound on $x^2 + y^2$ is an expression of the
Cauchy-Schwarz inequality.  For each set of values of
$(u,v,x,y)$, the SCS $|\underline{z},\underline{w}>$ form 
an orbit under $SU(3)$ action.  On each orbit we can
choose a convenient representative point $\left(\underline{z}
^{(0)}, \underline{w}^{(0)}\right)$, with any other point
$(\underline{z}, \underline{w})$ on the orbit arising from 
$\left(\underline{z}^{(0)}, \underline{w}^{(0)}\right)$
via suitable $SU(3)$ action as $\left(A\underline{z}^{(0)},
A^*\underline{w}^{(0)}\right)$.   The complete list of
orbits, representative points, stability subgroups
$H\left(\underline{z}^{(0)}, \underline{w}^{(0)}\right)
\subset SU(3)$ and orbit dimensions are as follows (with
$x,y$ omitted when irrelevant):
     \begin{eqnarray}
     a)~~\vartheta_1&=&\{u,v| u=v=0\}, 
     (\underline{z}^{(0)},\underline{w}^{(0)})=(\underline{0},\underline{0}), 
      H=SU(3),\mbox{dimension}\;0;\nonumber\\
     b)~~\vartheta_2&=&\{u,v| u>0,v=0\} , \underline{z}^{(0)} = u(1,0,0)^T,
     \underline{w}^{(0)}=\underline{0}, H=SU(2),\;
     \mbox{dimension}\;5;\nonumber\\
     c)~~\vartheta_3&=&\{u,v|u=0, v>0\} , \underline{z}^{(0)}=\underline{0},
     \underline{w}^{(0)}=v(0,1,0)^T, H=SU(2),\;
     \mbox{dimension}\;5;\nonumber\\
     d)~~\vartheta_4&=&\{u,v,x,y|u,v>0 ,0\leq x^2+y^2<1\} ,\nonumber\\
      && \underline{z}^{(0)} =
     u(1,0,0)^T, \underline{w}^{(0)} = 
     v\left(x+iy, \sqrt{1-x^2-y^2},0\right)^T,
      H=\{e\},\;\mbox{dimension}\;8;\nonumber\\
     e)~~\vartheta_5&=& \{u,v,x,y|u,v>0,\;x^2+y^2=1\},\nonumber\\
    && \underline{z}^{(0)} = u(0,0,1)^T,\;
     \underline{w}^{(0)} = v(x+iy)(0,0,1)^T,
     H=SU(2),\;\mbox{dimension}\;5.
     \label{79}
     \end{eqnarray}

\noindent
We add some comments: Class (a) comprises just the Fock vacuum
$|\underline{0},\underline{0}>$, invariant under $SU(3)$ and
forming a trivial orbit by itself.  Classes (b) and (c) form
collections of orbits with one of $\underline{z}$ and 
$\underline{w}$ vanishing identically, so these are simply SCS
for systems of three oscillators.  Class (d) is a four parameter 
family consisting of generic orbits.  Each orbit in this Class is 
eight dimensional and is essentially the $SU(3)$ group manifold.
Class (e) is a limiting form, as $x^2+y^2\rightarrow 1$, of
Class (d); in these orbits, $\underline{w}$  is a complex
multiple of $\underline{z}^*$.  However the limit is a singular
one, as is evident from the rise in the dimension of $H$
from zero to three, and the drop in orbit dimension from eight
to five.  This is why we have listed Class (e) separately.
Moreover, the 
representative point $(z^{(0)}, w^{(0)})$ in this class has been chosen 
so that the stability group $SU(2)$ acts on dimensions 1 and 2, thus 
coinciding with the subgroup relevant for the canonical basis $(\ref{84})$.
Disregarding Class (a), and recalling that ${\cal C}^6$ is of
real dimension 12, we see that Classes (b), (c), (d), (e) are non
overlapping regions in ${\cal C}^6$ of real dimensions 6, 6, 12
and 8 respectively.  Thus almost all of ${\cal C}^6$ is covered by 
orbits of Class (d).

Based on this orbit structure, we now express the Klauder
resolution of the identity, eqn.$(\ref{75})$, in a manner similar to
eqn.$(\ref{56})$, namely as an integration over the $SU(3)$ 
manifold followed by an integration over the invariants $(\ref{78})$.
(The difference compared to the case of two degrees of freedom 
is that here we integrate over the whole of $SU(3)$, not just
over a coset space such as $SU(2)/U(1)=S^2$  in eqn.$(\ref{56}))$.
In this process we can limit ourselves to Class (d) orbits
which are generic, as long as we do not at any later stage alter 
the integrand of eqn.$(\ref{75})$ by inserting a Dirac delta function
with support in one of the exceptional orbits in eqn.$(\ref{79})$.
To obtain a general pair $(\underline{z},\underline{w})$ from
$\left(\underline{z}^{(0)},\underline{w}^{(0)}\right)$ in
eqn.(4.16) Class (d), we need to parametrise (almost all) 
elements of $SU(3)$ in a convenient manner.  Here we use the fact
that, except on a set of vanishing measure, each $A\in SU(3)$
is uniquely determined by a pair $\left({\underline {\hat \eta}},
{\underline {\hat\zeta}}\right)$,
where $\underline{\hat{\eta}}$ is a complex three-component
unit vector and $\underline{\hat{\zeta}}$ is a complex two-
component unit vector\cite{28}:
     \begin{eqnarray}
     \underline{\hat{\eta}} &=&\left(\hat{\eta}_1,\hat{\eta}_2,
     \hat{\eta}_3\right)^T\;,\;\underline{\hat{\zeta}}  =
     \left(\hat{\zeta}_2,\hat{\zeta}_3\right)^T,\nonumber\\
     \underline{\hat{\eta}}^{\dag}\underline{\hat{\eta}}&=&
     \underline{\hat{\zeta}}^{\dag}\underline{\hat{\zeta}} = 1.
     \end{eqnarray}

\noindent
Then we have:
     \begin{eqnarray}
     A\in\;SU(3)&\Longleftrightarrow &A=A\left(\underline{\hat{\eta}},
     \underline{\hat{\zeta}}\right) =
     A_3\left(\underline{\hat{\eta}}\right)\;
     A_2\left(\underline{\hat{\zeta}}\right),\nonumber\\
     &&\nonumber\\
     A_3(\hat{\eta})&=&\left(\matrix{
      \hat{\eta}_1&\rho_1&0\cr \hat{\eta}_2&-\hat{\eta}_2\hat{\eta}^*_1/
     \rho_1&\hat{\eta}^*_3/\rho_1\cr
     \hat{\eta}_3&-\hat{\eta}_3\hat{\eta}^*_1/\rho_1&
      -\hat{\eta}^*_2/\rho_1}\right)\;\in\; SU(3),\nonumber\\
     &&\nonumber\\
     \rho_1&=&\left(1-|\hat{\eta}_1|^2\right)^{1/2} ;\nonumber\\
     &&\nonumber\\
     A_2\left(\underline{\hat{\zeta}}\right) &=& 
     \left(\matrix{
     1&0&0\cr 0&\hat{\zeta}_2&-\hat{\zeta}^*_3\cr
     0&\hat{\zeta}_3&\hat{\zeta}^*_2}\right)\;
     \in\;SU(2)\subset SU(3) .
     \end{eqnarray}

\noindent
For each $\underline{\hat{\eta}}$ (provided $|\hat{\eta}_1|< 1$),
$A_3\left(\underline{\hat{\eta}}\right)$ is a particular $SU(3)$
element completely determined by its first column which is 
$\underline{\hat{\eta}}$; and for each $\underline{\hat{\zeta}},
A_2\left(\underline{\hat{\zeta}}\right)$ is an element in the
$SU(2)$ subgroup leaving $\underline{z}^{(0)}$ invariant.
We can picture $\underline{\hat{\eta}}$ and $\underline{\hat{\zeta}}$
as representing points on $S^5\subset{\cal R}^6$ and $S^3\subset
{\cal R}^4$ respectively.  Then the normalised invariant volume
element on $SU(3)$ is a numerical factor times the product 
of the solid angle elements on $S^5$ and $S^3$:
     \begin{eqnarray}
     dA\left(\underline{\hat{\eta}},\underline{\hat{\zeta}}
     \right)
     &=&(2\pi^5)^{-1}\;d\Omega_5\left(\underline{\hat{\eta}}\right)
     \;d\Omega_3\left(\underline{\hat{\zeta}}\right),\nonumber\\
     \int\limits_{SU(3)}\;dA&=& 1.
     \end{eqnarray}

\noindent
The expressions for $(\underline{z},\underline{w})$ in terms of 
$A\left(\underline{\hat{\eta}},\underline{\hat{\zeta}}\right)$
and $\left(\underline{z}^{(0)},\underline{w}^{(0)}\right)$ are:
     \begin{eqnarray}
     \underline{z}&=&A\left(\underline{\hat{\eta}},\underline
     {\hat{\zeta}}\right)\;\underline{z}^{(0)}(u) =
     u \;\underline{\hat{\eta}},\nonumber\\
     \underline{w}&=&A\left(\underline{\hat{\eta}},
     \underline{\hat{\zeta}}\right)^*\;\underline{w}^{(0)}
     (v,x,y) = v\;A_3\left(\underline{\hat{\eta}}\right)^*
     \left(\matrix{
     x+iy &\;\;\cr
     \sqrt{1-x^2-y^2}&\hat{\zeta}^*_2\cr
     \sqrt{1-x^2-y^2}&\hat{\zeta}^*_3}\right).
     \end{eqnarray}

\noindent
These are the generalisations of eqn.$(\ref{51})$.  Straight forward 
computations of the Jacobians yield:
     \begin{eqnarray}
     \prod\limits^{3}_{j=1}\left(d^2z_j\;d^2w_j\right) =
     u^5 v^5(1-x^2-y^2) du\;dv\;dx\;dy\;d\Omega_5
     \left(\underline{\hat{\eta}}\right)\;d\Omega_3
     \left(\underline{\hat{\zeta}}\right).
    \label{84}
    \end{eqnarray}

\noindent
We can now rewrite the Klauder result $(\ref{75})$ as:
     \begin{eqnarray}
     \frac{2}{\pi}\int\limits^{\infty}_0\;u^5 du\int\limits
     ^{\infty}_0\;v^5\;dv&&\int\limits_{x^2+y^2\leq 1}\;
     (1-x^2-y^2) dx\;dy\int\limits_{SU(3)}\;dA\;{\cal U}(A)|
     \underline{z}^{(0)}(u), \underline{w}^{(0)} (v,x,y)
     \rangle \times\nonumber\\
     &&\nonumber\\ 
     &&\langle\underline{z}^{(0)}(u), \underline{w}^{(0)}
     (v,x,y)|{\cal U}(A)^{-1} = 1\;\mbox{on}\;{\cal H}.
     \label{85}
     \end{eqnarray}

\noindent
This is the analogue of (the initial form of) eqn.$(\ref{56})$.

In the spirit of eqns.$(\ref{27}, \ref{59})$ we can now consider
modifications of eqn.$(\ref{85})$ by including in the integrand a
function of the $SU(3)$ invariants.  Thus we define
     \begin{eqnarray}
     A(f)&=&\int\;\prod\limits^3_{j=1}\left(\frac{d^2z_j}
     {\pi}\;\frac{d^2w_j}{\pi}\right)\;f(u,v,x,y)\;
     |\underline{z},\underline{w}><\underline{z},\underline{w}|
    \nonumber\\
     &&\nonumber\\
     &=&\frac{2}{\pi}\;\int\limits^{\infty}_0\;u^5 du\int\limits
     ^{\infty}_0\;v^5dv\;\int\limits_{x^2+y^2\leq 1}
     (1-x^2-y^2) dx \;dy \;f(u,v,x,y)\;\int\limits_{SU(3)}\;
     dA\;\times\nonumber\\
     &&\nonumber\\
     &&{\cal U}(A)|\underline{z}^{(0)}(u),\underline{w}^{(0)}
     (v,x,y)><\underline{z}^{(0)}(u), \underline{w}^{(0)}
     (v,x,y)|{\cal U}(A)^{-1}.
     \label{86}
     \end{eqnarray}

\noindent
Such an operator definitely obeys
     \begin{eqnarray}
     {\cal U}(A)\;A(f) = A(f)\;{\cal U}(A),\;\mbox{all}\;A\in\;SU(3).
     \end{eqnarray}

\noindent
However, as long as $f(u,v,x,y)$ is nontrivial, the measure in 
eqn.$(\ref{86})$ is not the invariant one on the H-W group, we do
not have recourse to Schur lemma for the UIR of this group,
and $A(f)$ is not proportional to the identity operator on
${\cal H}$.  The presence of (infinite!) multiplicity in
the reduction of ${\cal U}(A)$ on ${\cal H}$ into UIR's of $SU(3)$
means furthermore that we do not immediately get for $A(f)$
a simple combination of $SU(3)$-invariant projections as we
did in eqns.$(\ref{59}, \ref{60})$ with $SU(2)$.

\noindent
\underline{The restriction to ${\cal H}_0$}

Now we limit ourselves to the SCS $|\underline{z},\underline{w}>$
belonging to ${\cal H}_0\subset {\cal H}$, as in this subspace
the multiplicity problem is avoided.  As noted following eqn.
$(\ref{74})$, the condition $\underline{z}^T\underline{w}=0$
ensures $|\underline{z},\underline{w}>\in\;{\cal H}_0$.  
This happens in Classes (a), (b), (c) of eqn.$(\ref{79})$ in a
trivial manner, and in Class (d) when $x=y=0$. The former 
can be disregarded as being sets of vanishing measure.

We deal first with vector level relations in ${\cal H}_0$,
then look at modifications of $A(f)$ in eqn.$(\ref{86})$.  We
begin with eqn.$(\ref{72})$.  For the highest weight states of
$SU(3)$ UIR's occurring there, we introduce a simpler notation:
     \begin{eqnarray}
      |p,q;\frac{1}{2}(p+q), \frac{1}{2}
     (p+q); \frac{1}{3}(p-q);\frac{1}{2}
      (p+q+3)\rangle&\equiv& |p,q;\frac{1}{2}
     (p+q), \frac{1}{2}(p+q); \frac{1}{3}(p-q)\rangle_0
      \nonumber\\
      &&\nonumber\\
      &&\in\;{\cal H}^{(p,q;0)}\subset {\cal H}_0.
      \end{eqnarray}

\noindent
We have omitted the $Sp(2,R)$ quantum number $m$ as it is
superfluous  within ${\cal H}_0$.  Then eqn.$(\ref{72})$ takes the form
     \[\underline{z}^{(0)}(u)= u(1,0,0)^T,\;\underline{w}
     ^{(0)}(v,0,0) = v(0,1,0)^T:\]
    \begin{eqnarray}
    |\underline{z}^{(0)}(u), \underline{w}^{(0)}(v,0,0)\rangle
    &=&e^{-\frac{1}{2}(u^2+v^2)}\sum\limits^{\infty}_{p,q=0}
    \frac{u^p \;v^q}{\sqrt{p!q!}}\nonumber\\
    &&\nonumber\\
     &&|p,q;\frac{1}{2}(p+q), \frac{1}{2}(p+q),
    \frac{1}{3}(p-q)\rangle_0\;\in\;{\cal H}_0.
    \label{89}
    \end{eqnarray}

\noindent
In place of eqn.$(\ref{68})$, the $SU(3)$ SCS within each UIR $(p,q)$
contained in ${\cal H}_0$ can be written as
     \begin{eqnarray}
     A\in\; SU(3) : |p,q;A\rangle_0 =
     {\cal U}(A)|p,q;\frac{1}{2}(p+q), \frac{1}{2}(p+q),
     \frac{1}{3}(p-q)\rangle_0\;\in\;{\cal H}^{(p,q;0)}
     \label{90}
     \end{eqnarray}

\noindent
Applying ${\cal U}(A)$ for general $A\in\;SU(3)$ to both sides of eqn.$
(\ref{89})$ we get a result linking those H-W SCS that lie in ${\cal H}_0$,
and the $SU(3)$ SCS $(\ref{90})$ within each UIR $(p,q)$ in 
${\cal H}_0$:
    \begin{eqnarray}
    A\in\;SU(3), \underline{z}&=& A\underline{z}^{(0)}(u),
    \underline{w}=A^*\underline{w}^{(0)}(v,0,0):\nonumber\\
    &&\nonumber\\
    |\underline{z},\underline{w}\rangle &=&
    e^{-\frac{1}{2}(u^2+v^2)}\sum\limits^{\infty}
    _{p,q=0} \frac{u^p\;v^q}{\sqrt{p!q!}}\;
    |p,q;A\rangle_0\;\in\;{\cal H}_0.
    \label{91}
    \end{eqnarray}

\noindent
This is the $SU(3)$ analogue to the $SU(2)$ relation
$(\ref{55})$.

Now we turn to the operator $A(f)$ in eqn.$(\ref{86})$ and
make the choice 
     \begin{equation}
     f(u,v,x,y) = f_0(u,v)\;\delta(x)\;\delta(y).
     \end{equation}

\noindent
This leads to
     \begin{eqnarray}
     A(f_0) &=&\frac{2}{\pi}\int\limits^{\infty}_0\;
     u^5 du\;\int\limits^{\infty}_0\;v^5 dv\;f_0(u,v)\;
     \int\limits_{SU(3)}\;dA\;{\cal U}(A)\nonumber\\
     &&\nonumber\\
     &&|\underline{z}^{(0)}(u),\underline{w}^{(0)}(v,0,0)
     \rangle\langle\underline{z}^{(0)}(u),\underline{w}
      ^{(0)} (v,0,0) \;|{\cal U}(A)^{-1}.
    \label{93}
    \end{eqnarray}

\noindent
Such an operator obeys the following:
     \begin{eqnarray}
     \psi\in {\cal H}_0^{\bot}&:&A(f_0)\psi=0;\nonumber\\
     \psi\in {\cal H}_0&:& A(f_0) \psi\in{\cal H}_0;\nonumber\\
     A\in\;SU(3)&:& {\cal U}(A)\;A(f_0) = A(f_0)\;{\cal U}(A).
     \end{eqnarray}

\noindent
Therefore $A(f_0)$ must be a linear combination of the
projection operators $P^{(p,q;0)}$  onto the subspaces
${\cal H}^{(p,q;0)}\subset {\cal H}_0$; it is here that we
exploit the multiplicity-free reduction of the $SU(3)$ UR
${\cal D}_0$ on ${\cal H}_0$.  To get $A(f_0)$ explicitly,
we use the following immediate consequences of Schur
lemma applied to $SU(3)$, the multiplicity-free nature of
${\cal D}_0$, and the orthogonality of inequivalent UIR's:
     \begin{eqnarray}
     \int\limits_{SU(3)} dA\;|p,q;A\rangle_0\;_0\langle 
     p^{\prime},q^{\prime}; A| = \delta_{p^{\prime} p}
     \delta_{q^{\prime}q}\;P^{(p,q;0)}\big/
     d(p,q) .
     \label{95}
     \end{eqnarray}

\noindent
Then a combination of eqns.$(\ref{93}, \ref{89}, \ref{90}, \ref{95})$ 
immediately gives:
     \begin{eqnarray}
     A(f_0)&=&\sum\limits^{\infty}_{p,q=0}\;C(f_0;p,q)\;
     P^{(p,q;0)},\nonumber\\
     C(f_0;p,q)&=& \{p! q!\;d(p,q)\}^{-1}\cdot
     \frac{2}{\pi}\cdot \int\limits^{\infty}_0
     \;u^5 du\int\limits^{\infty}_0\;v^5 dv\;f_0
     (u,v) u^{2p}v^{2q}e^{-(u^2+v^2)}.
     \label{96}
     \end{eqnarray}

\noindent
This is an $SU(3)$ analogue of the $SU(2)$ result $(\ref{59})$, but
it is valid only after the restriction to ${\cal H}_0$.
On account of the freedom still remaining in eqns.$(\ref{93},\ref{96})$ in 
the choice of the function $f_0(u,v)$, we see that the H-W
SCS occurring there are overcomplete in ${\cal H}_0$.
If we wish to limit ourselves to an exact ``$SU(3)$-
worth'' of H-W SCS within ${\cal H}_0$, then we have the
analogue to eqn.$(\ref{60})$:
     \begin{eqnarray}
     f_0(u,v) &=&\delta(u-u_0)\;\delta(v-v_0):\nonumber\\
     A(f_0)&=&\int\;\prod\limits^{3}_{j=1}\left(
     \frac{d^2z_j}{\pi}\;\frac{d^2w_j}{\pi}\right)
      \delta(u-u_0)\delta(v-v_0)\delta(x)
     \delta(y)|\underline{z},\underline{w}\rangle\langle
      \underline{z},\underline{w}|\nonumber\\
     &=&e^{-\left(u^2_0 +v^2_0\right)}\cdot
     \frac{2}{\pi}\;\sum\limits^{\infty}_{p,q=0}\;
     u^{2p+5}_0 v_0^{2q+5} P^{(p,q;0)}/
     p!q!d(p,q).
     \label{97}
     \end{eqnarray}

\noindent
The point to be emphasised is how far this result departs from
being the identity operator in ${\cal H}_0$, leave alone in 
${\cal H}$, but understandably so.

\noindent
\underline{Description in ${\cal H}^{({\rm ind})}$}

As recalled in Section II, and established in detail in I, the
multiplicity-free UR ${\cal D}_0$ of $SU(3)$ on ${\cal H}_0$
is equivalent to an induced UR ${\cal D}^{({\rm ind})}$ of
$SU(3)$, namely the one arising from the trivial representation
of an $SU(2)$ subgroup of $SU(3)$.  The isomorphism between
${\cal H}_0$ and ${\cal H}^{({\rm ind})}$ carrying
${\cal D}^{({\rm ind})}$, consistent with the two group actions,
is given in eqn.$(\ref{17})$.  It is of interest to see what wave 
functions $\psi(\underline{\xi})\in\;{\cal H}^{({\rm ind})}$
one obtains for the various vectors in ${\cal H}_0$ that have
played a role earlier in this Section.  We now give these wave
functions and comment briefly on them.

For  the highest weight state in the $SU(3)$ UIR $(p,q)$
on ${\cal H}^{(p,q;0)}$, and the associated $SU(3)$
SCS, we find the following wavefunctions in 
${\cal H}^{({\rm ind})}$:
     
     \[|p,q;\frac{1}{2}(p+q), \frac{1}{2}(p+q), \frac{1}{3}(p-q)
     \rangle_0\longrightarrow \sqrt{\frac{(p+q+2)!}{p!q!}}\;
     (\xi_1)^p \left(\xi^*_2\right)^q ;\]
     \begin{eqnarray}
     |p,q;A\rangle_0&=& {\cal U}(A)|p,q;\frac{1}{2}(p+q), \frac{1}{2}
     (p+q), \frac{1}{3}(p-q)\rangle_0\longrightarrow\nonumber\\
     &&\nonumber\\
     &&\sqrt{\frac{(p+q+2)!}{p!q!}}\;
     \left(A^*_{j1} \xi_j\right)^p
     \left(A_{k2} \xi^*_k\right)^q .
     \end{eqnarray}

\noindent
For the H-W SCS in ${\cal H}_0$ generating these states
within each UIR we have:
     \begin{eqnarray*}
     |\underline{z}^{(0)}(u), \underline{w}^{(0)}(v,0,0)\rangle
     \longrightarrow e^{-\frac{1}{2}(u^2+v^2)}\sum\limits^{\infty}
     _{p,q=0}\;\sqrt{(p+q+2)!}\;\frac{(u\xi_1)^p}{p!}\;
     \frac{\left(v \xi^*_2\right)^q}{q!} ;
     \end{eqnarray*}
     \begin{eqnarray}
     |\underline{z},\underline{w}\rangle &=&{\cal U}(A)|
     \underline{z}^{(0}(u), \underline{w}^{(0)}
     (v,0,0)\rangle\longrightarrow\nonumber\\
      &&\nonumber\\
     &&e^{-\frac{1}{2}(u^2+v^2)}\sum\limits^{\infty}
     _{p,q=0}\;\sqrt{(p+q+2)!}\;
     \frac{\left(u A^*_{j1}\xi_j\right)^p}{p!}\;
     \frac{\left(v A_{k2}\xi^*_k\right)^q}{q!}.
     \end{eqnarray}

\noindent
The principal comment we may make is that these particular H-W 
SCS do not have wave functions in ${\cal H}^{({\rm ind})}$ in
the form of any simple expressions involving exponential functions.
The reason for this can be traced to the factorial in eqn.$(\ref{17})$
as compared to eqn.$(\ref{14}a)$.  Another way of understanding
this situation is to realise that ${\cal H}_0$ (and so 
${\cal H}^{({\rm ind})}$ as well) is too small to carry a 
representation  of the H-W system used in the Schwinger $SU(3)$ 
construction; in addition the argument $\underline{\xi}$
in $\psi(\underline{\xi})$ is a complex unit vector in three
dimensions rather than a variable in all of ${\cal C}^3$.

\section{General eigenspaces ${\cal H}_{\kappa}$ of $K_-$}

The subspace ${\cal H}_0\subset {\cal H}$ carrying the
multiplicity-free UR ${\cal D}_0$ of $SU(3)$, the focus of 
analysis in the preceding Section, is spanned by those
H-W SCS $|\underline{z},\underline{w}>$ for which 
$\underline{z}^T\underline{w}=0$, and belonging to a
particular collection of orbits under Class (d) of eqn.$(\ref{79})$:
     \begin{eqnarray}
     {\cal H}_0 = Sp\{|\underline{z}, \underline{w}>|
      \underline{z}, \underline{w} \in {\cal C}^3,\;
      \underline{z}^T\underline{w}=0\}.
      \label{100}
      \end{eqnarray}

\noindent
As noted earlier, these SCS are actually over  complete
within ${\cal H}_0$.  Since, by eqn.$(\ref{74})$, $\underline{z}^T
\underline{w}$ is the eigenvalue of the $SU(3)$ invariant
$Sp(2,R)$ lowering operator $K_-$, this means that
${\cal H}_0$ is spanned by those H-W SCS that are
eigenvectors of $K_-$ with eigenvalue zero.  Moreover,
eqns.$(\ref{89},\ref{91})$ show that these H-W SCS are directly 
connected to the $SU(3)$ SCS within each $SU(3)$ UIR
$(p,q)$, carried by ${\cal H}^{(p,q;0)}\subset 
{\cal H}_0$.

It now turns out that a somewhat similar situation exists 
involving eigenvectors of $K_-$ corresponding to nonzero
eigenvalues as well, but with one major difference: we
encounter certain specific $SU(3)$ GCS systems.  This also 
connects up with a certain class of coherent states within
the UIR's $D_k^{(+)}$ of $Sp(2,R)$.  We analyse these matters
in this Section.It turns out that H-W SCS 
of both Classes (d) and (e) are involved.

We begin by generalising eqn.$(\ref{100})$ and defining a subspace
${\cal H}_{\kappa}\subset {\cal H}$, for any complex number
$\kappa$, as consisting of eigenvectors of $K_-$ with
eigenvalue $\kappa$; equally well it is the span of all
those H-W SCS which obey this condition:
     \begin{eqnarray}
     {\cal H}_{\kappa}&=&\{|\psi> \in {\cal H}| K_-|\psi>
     =\kappa|\psi>\}\nonumber\\
      &=&Sp\{|\underline{z},\underline{w}>
      |\underline{z},\underline{w} \in {\cal C}^3,\;
      \underline{z}^T \underline{w} = \kappa\}\subset
      {\cal H} .
      \label{101}
      \end{eqnarray}

\noindent
These H-W SCS comprise a particular subset of Class (d)
orbits in eqn.$(\ref{79})$; for $\kappa=0$ we get back 
${\cal H}_0$.  It is important to remark that even
though $\kappa$ varies over a continuum, each
${\cal H}_{\kappa}$ consists of bona fide (ie., normalisable)
vectors in ${\cal H}$; and for $\kappa^{\prime}\neq \kappa,
\;{\cal H}_{\kappa^{\prime}}$ and ${\cal H}_{\kappa}$ are
not mutually orthogonal.  As in the case of the oscillator
annihilation operator, these are consequences of $K_-$
being non hermitian.  Since $K_-$ is $SU(3)$ invariant,
each ${\cal H}_{\kappa}$ is $SU(3)$ invariant as well:
     \begin{equation}
     A\in\;SU(3),\;|\psi>\in\;{\cal H}_{\kappa}\Longrightarrow
     {\cal U}(A)|\psi>\in\;{\cal H}_{\kappa}.
     \end{equation}

\noindent
Therefore the UR ${\cal U}(A)$ of $SU(3)$ on ${\cal H}$, when 
restricted to ${\cal H}_{\kappa}$ leads to a UR ${\cal D}_{\kappa}$
acting on ${\cal H}_{\kappa}$.  We will see that this UR
contains each UIR $(p,q)$ exactly once, just like ${\cal D}_0$
on ${\cal H}_0$.  Thus it is also multiplicity-free and complete.

To exhibit these properties, we first recall the  
construction of eigenvectors of $K_-$ in any discrete class 
UIR $D^{(+)}_k$ of $Sp(2,R)$\cite{29}.  (Though the following results are 
valid for all real $k>0$, we require only the cases $k=3/2,2,5/2
\ldots$).  As in eqn.(I.3.24,25), denote the eigenvectors of $J_0$
in $D^{(+)}_k$ by $|k,m>$.  Then we have the well-known results:
     \begin{mathletters}
     \begin{eqnarray}
     |k,\kappa> &=&\left(_0 F_1(2k;|\kappa |^2)\right)^{-1/2}
     \sum\limits^{\infty}_{m=k} (\Gamma(2k)/(m-k)!
     \Gamma(m+k))^{1/2}\;
      \kappa^{m-k}|k,m>,\nonumber\\
      K_-|k,\kappa>&=& \kappa |k,\kappa>,\;\;\kappa\in{\cal C};\\
     <k,\kappa^{\prime}|k,\kappa>&=&_0F_1(2k;\kappa^{\prime^{*}}
     \kappa)
     \bigg/\left\{_0F_1(2k;|\kappa^{\prime}|^2)_0F_1
     (2k; |\kappa |^2)\right\}^{1/2} ;\\
      &&\int\limits_{{\cal C}}\;\frac{d^2\kappa}{\pi}\;
     \sigma(|\kappa|^2) |k,\kappa><k,\kappa| = 1, \nonumber\\
      \sigma(|\kappa|^2)&=&\frac{2}{\Gamma(2k)}
      \;_0F_1(2k; |\kappa|^2) |\kappa|^{2k-1}
       K_{\frac{1}{2}-k}(2|\kappa|), 
       \end{eqnarray}
       \label{103}
      \end{mathletters}
where $K_\nu (z)$ denotes modified Bessel function of the third kind. 
\noindent
(For simplicity the $k$-dependence of the weight function
$\sigma$ is omitted).  We note that even though these states 
$|k,\kappa>$ within $D^{(+)}_k$ do not form an $Sp(2,R)$ orbit, 
they do furnish a Klauder-type resolution of the identity.

We now exploit this construction in the present context.
We begin with two facts: (a) the vectors $|p,q; IMY;m>$, 
as all labels vary, form an orthonormal basis for the total 
Hilbert space ${\cal H}$; (b) if we keep $p,q, IMY$ fixed and 
allow only $m$ to vary, we get an orthonormal basis for a 
subspace carrying just the UIR $D^{(+)}_k$ of $Sp(2,R)$.  
Therefore, in view of the construction $(\ref{103})$, within each such
subspace we can define and have:
     \begin{eqnarray}
     |p,q; IMY\rangle_{\kappa} &=&
     \left\{_0F_1(2k;|\kappa|^2)\right\}^{-1/2}\sum
     \limits^{\infty}_{m=k} ((2k-1)!/(m-k)!(m+k-1)!)^{1/2}
     \times\nonumber\\
     &&\kappa^{m-k} |p,q; IMY; m\rangle ,\nonumber\\
      K_- |p,q; IMY\rangle_{\kappa}&=&
     \kappa |p,q; IMY\rangle_{\kappa} ;\nonumber\\
     _{\kappa^{\prime}}\langle p^{\prime},q^{\prime};
     I^{\prime}M^{\prime}Y^{\prime} | p,q;IMY\rangle_{\kappa}&=&
     \delta_{p^{\prime}p}\delta_{q^{\prime}q}\delta_{I^{\prime}I}
     \delta_{M^{\prime}M}\delta_{Y^{\prime}Y}\times\nonumber\\
     _0F_1\left(2k; \kappa^{\prime^{*}}\kappa\right)&\bigg/ &
     \left\{_0F_1(2k;|\kappa^{\prime}|^2)
     _0F_1(2k;|\kappa|^2)\right\}^{1/2}.
     \label{104}
     \end{eqnarray}

\noindent
(For fixed $p,q,IMY$ we also have a resolution of the appropriate
identity in the form of eqn.$(\ref{103}c)$, but we omit it).  For 
$\kappa =0$ we recover the orthonormal basis for ${\cal H}_0$.
However for $\kappa\neq 0$, these vectors are not eigenvectors
of the total $a$-type and $b$-type number operators $\hat{N}^{(a)},
\hat{N}^{(b)}$.  It is now evident that if we keep $\kappa$
fixed, allow $pq IMY$ to vary, and recall that the range 
${\cal C}$ of $\kappa$ is $k$-independent, we get an orthonormal
basis for ${\cal H}_{\kappa}$:
     
     \[{\cal H}_{\kappa} = Sp\{|p,q; IMY\rangle_{\kappa} |\kappa
     \;\mbox{fixed},\; pq IMY\;\mbox{varying}\},\] 
     \begin{equation}
     _{\kappa}\langle  p^{\prime}, q^{\prime}; I^{\prime}
     M^{\prime} Y^{\prime} | p,q; IMY\rangle_{\kappa} =
     \delta_{p^{\prime}p}\delta_{q^{\prime}q}\delta_{I^{\prime}I}
     \delta_{M^{\prime}M}\delta_{Y^{\prime}Y}.
     \label{105}
     \end{equation}

\noindent
It is also clear that each UIR $(p,q)$ of $SU(3)$, carried by the
$d(p,q)$ vectors $|p,q; IMY\rangle_{\kappa}\in\;{\cal H}_{\kappa}$
as $IMY$ alone vary, appears exactly once in ${\cal H}_{\kappa}$.
In other words, ${\cal D}_{\kappa}$ is multiplicity-free.
In eqns.$(\ref{101},\ref{105})$ we have three equally good ways of identifying
the subspace ${\cal H}_{\kappa}\subset {\cal H}$. 

We next relate the orthonormal basis vectors $(\ref{105})$ for
${\cal H}_{\kappa}$ to the corresponding ones for ${\cal H}_0$
in eqn.$(\ref{12})$, in a compact manner.  For this we use eqn.
(I.3.25b) valid within each UIR $D^{(+)}_k$ of $Sp(2,R)$,
along with $K_+=\underline{{\hat a}}^{\dag}\cdot \underline{{\hat b}}
^{\dag}$:
     \begin{eqnarray}
     |p,q;IMY;m\rangle &=&
     ((2k-1)!/(m-k)!(m+k-1)!)^{1/2}\left(\underline{{\hat a}}^{\dag}\cdot
     \underline{{\hat b}}^{\dag}\right)^{m-k}|p,q; IMY;k\rangle;\nonumber\\
      &&\nonumber\\
     |p,q; IMY\rangle_{\kappa}&=&\left\{_0F_1(2k;|\kappa|^2)
     \right\}^{-1/2} \sum\limits^{\infty}_{m=k}
     \frac{(2k-1)!}{(m-k)!(m+k-1)!}\times\nonumber\\
     &&\nonumber\\
     &&\left(\kappa \underline{{\hat a}}^{\dag}\cdot\underline{{\hat b}}^{\dag}
     \right)^{m-k} |p,q; IMY; k\rangle\nonumber\\
      &&\nonumber\\
     &=&A^{\dag}_{k,\kappa} |p,q; IMY\rangle_0,\nonumber\\
      &&\nonumber\\
     A^{\dag}_{k,\kappa}&=&\left\{_0F_1(2k; |\kappa|^2)\right\}
     ^{-1/2} \sum\limits^{\infty}_{m=k}
     \frac{(2k-1)!}{(m-k)!(m+k-1)!}
     \;\left(\kappa \underline{{\hat a}}^{\dag}\cdot\underline{{\hat b}}^{\dag}
     \right)^{m-k}\nonumber\\
     &&\nonumber\\
     &=& _0F_1\left(2k; \kappa \underline{{\hat a}}^{\dag}\cdot 
     \underline{{\hat b}}
     ^{\dag}\right) \bigg/\left\{_0F_1(2k; |\kappa|^2)\right\}^{1/2}.
     \label{106}
     \end{eqnarray}

\noindent
It is important to notice that there is a dependence on 
$k=\frac{1}{2}(p+q+3)$ in the operator $A^{\dag}_{k,\kappa}$;
so the basis vectors $|p,q;IMY\rangle_{\kappa}$ for
${\cal H}_{\kappa}$ do not arise from the basis vectors
$|p,q;IMY\rangle_0$ for ${\cal H}_0$  by application of a
single operator dependent on $\kappa$ alone.  In spite
of this, we will see below the usefulness of the
connection $(\ref{106})$.

We now obtain an expansion of the H-W SCS $|\underline{z},
\underline{w}>$ with $\underline{z}^T\underline{w}=\kappa$,
in the orthonormal basis $(\ref{105})$ for ${\cal H}_{\kappa}$.
Thus we seek analogues to eqns.$(\ref{89},\ref{91})$, as well as to
eqns.$(\ref{90},\ref{95})$, in the case of ${\cal H}_0$.  Given
$|\underline{z},\underline{w}\rangle \in {\cal H}_{\kappa}$,
by a suitable $SU(3)$ transformation we can relate it to a
standard state $|\underline{z}^{(0)}(u), \underline{w}^{(0)}
(v,x,y)\rangle$ on its orbit.  We parametrise the latter as in 
eqn.$(\ref{79})$ Class (d)( We are assuming here for definiteness that 
$x^2+y^2 < 1$, the possibility $x^2+y^2=1$ which is of vanishing 
measure being handled in the next Section):
     \begin{eqnarray}
     \underline{z}^{(0)}(u)&=& u(1,0,0)^T,\nonumber\\
     \underline{w}^{(0)}(v,x,y) &=& v\left(x+iy,
     \sqrt{1-x^2-y^2},0\right)^T,\nonumber\\
     uv(x+iy)&=&\kappa .
     \end{eqnarray}

\noindent
We develop first the replacement for eqn.$(\ref{89})$.  The
point of interest is to see which vector within each
UIR $(p,q)$ in ${\cal D}_{\kappa}$ appears, in place of
the higher weight vector present in eqn.$(\ref{89})$.  Thanks
to eqn.$(\ref{106})$, the relevant overlap simplifies to a
calculation in ${\cal H}_0$:
     \begin{eqnarray*}
     _{\kappa}\langle p,q; IMY|\underline{z}^{(0)}(u),
      \underline{w}^{(0)}(v,x,y)
     \rangle = _0\langle p,q; IMY| A_{k,\kappa}
     |\underline{z}^{(0)}(u), \underline{w}^{(0)}
     (v,x,y)\rangle
     \end{eqnarray*}
     \begin{eqnarray}
     =\left\{_0F_1(2k;|\kappa|^2)\right\}^{1/2}
     \langle p,q; IMY; k|\underline{z}^{(0)}(u),
     \underline{w}^{(0)}(v,x,y)\rangle .
     \label{108}
     \end{eqnarray}

\noindent
Here the bra vector, in ${\cal H}_0$, is an eigenvector of
$\hat{N}^{(a)}, \hat{N}^{(b)}$ with eigenvalues $p,q$ 
respectively.  This leads to further simplification:
     \begin{eqnarray*}
     \langle p,q; IMY;k| \underline{z}^{(0)}(u), 
     \underline{w}^{(0)}(v,x,y)\rangle =
     e^{-\frac{1}{2}(u^2+v^2)} \frac{u^p}{p!}\;
      \frac{v^q}{q!}\times
     \end{eqnarray*}
     \begin{eqnarray}
     \langle p,q;IMY; k| \left({\hat a}^{\dag}_1\right)^p
     \left((x+iy)\;{\hat b}_1^{\dag} +\sqrt{1-x^2-y^2}\; {\hat b}_2^{\dag}
     \right)^q |\underline{0},\underline{0}\rangle .
     \label{109}
     \end{eqnarray}

\noindent
The ket vector here has hypercharge $\frac{1}{3}(p-q)$, as
does the highest weight state in $(p,q)$, so this overlap
is nonzero only if $Y=\frac{1}{3}(p-q)$.  This then
determines the possible values of $I$:
     \begin{eqnarray}
     I&=& I_0, I_0-1,\ldots, \frac{1}{2}|p-q| ,\nonumber\\
     I_0&=& \frac{1}{2}(p+q).
     \label{110}
     \end{eqnarray}

\noindent
Notice that $I_0$ is the highest possible value of $I$ in the 
UIR $(p,q)$.  For the bra vector in eqn.$(\ref{109})$ we have the
explicit expression (eqn.(I.A.9)):
     \begin{eqnarray}
     \langle p,q; IMY;k|&=& {\cal N}_{pqIY}((I+M)!(I-M)!/
     2I!)^{1/2} \sum\limits^{(p-r,q-s)_<}_{n=0}
     \sum\limits^{I-M}_{L=0}\times\nonumber\\
      &&\nonumber\\
     &&\frac{(-1)^{n+I-M-L}}{(r+s+n+1)!}\;\langle\underline{0},
     \underline{0}|\;
     \frac{({\hat a}_{\alpha}{\hat b}_{\alpha})^n}{n!} \;
     \frac{{\hat a}_{1}^{r-L}}{(r-L)!}\;\frac{{\hat a}_2^L}{L!}
     \times \nonumber\\
      &&\nonumber\\
     &&\frac{{\hat b}_{1}^{I-M-L}}{(I-M-L)!}\;
     \frac{{\hat b}_{2}^{s-I+M+L}}{(s-I+M+L)!}\;
     \frac{{\hat a}_{3}^{p-r-n}}{(p-r-n)!}\;
     \frac{{\hat b}_{3}^{q-s-n}}{(q-s-n)!}\;,\nonumber\\
      &&\nonumber\\
     {\hat a}_{\alpha}\;{\hat b}_{\alpha}&=& {\hat a}_1\;{\hat b}_1 
     + {\hat a}_2\;{\hat b}_2\;,\nonumber\\
     &&\nonumber\\
      {\cal N}_{pqIY}&=& \{r!s!(r+s+1)! (p-r)! (q-s)! (p+s+1)!
     (q+r+1)!/(p+q+1)!\}^{1/2}\;,\nonumber\\
      &&\nonumber\\
     r&=& I+\frac{Y}{2} + \frac{1}{3} (p-q),\; S=I-\frac{Y}{2} +
     \frac{1}{3} (q-p) .
     \label{111}
     \end{eqnarray}

\noindent
Use of this in eqn.$(\ref{109})$ leads to further simplifications.
The condition $Y=\frac{1}{3}(p-q)$ gives:
     \begin{eqnarray}
     r=I&+& M_0,\; s=I-M_0,\nonumber\\
     p-r&=& q-s = I_0 -I,\nonumber\\
     M_0&=& \frac{1}{2}(p-q) .
     \end{eqnarray}

\noindent
Then, in the sums over $n$ and $L$ in eqn.$(\ref{111})$, only the
terms $n=p-r=q-s$ and $L=0$ survive.  Using all this, the
scalar product in eqn.$(\ref{108})$ can be explicitly computed:
     \begin{eqnarray*}
     _{\kappa}\langle p,q;IMY|\underline{z}^{(0)}(u),
     \underline{w}^{(0)}(v,x,y)\rangle = \left\{
     _0F_1(2k;|\kappa|^2)\right\}^{1/2}\cdot
     e^{-\frac{1}{2}(u^2+v^2)}\cdot
     u^p\;v^q\;\times
     \end{eqnarray*}

     \begin{eqnarray*}
     \delta_{Y,\frac{1}{3}(p-q)}\;
     \frac{(-1)^{I_0-M}}{(M-M_0)!}\
     \{(2I+1)(I+M)! (I-M_0)!/(2I_0+1)! (I-M)!(I+M_0)!\}^{1/2}\times
     \end{eqnarray*}
     \begin{eqnarray*}
     (x+iy)^{I_0-M} (1-x^2 -y^2)^{\frac{1}{2}(M-M_0)},
     \end{eqnarray*}
     \begin{equation}
      u\;v(x+i\;y) = \kappa .
     \label{113}
     \end{equation}

\noindent
We see that, provided $Y=\frac{1}{3}(p-q)$ and $M\geq M_0$, this 
overlap is nonzero for all values of $I$ in the range $(\ref{110})$.
This shows how far the projection of $|\underline{z}^{(0)}(u),
\underline{w}^{(0)}(v,x,y)\rangle$ onto the subspace of
${\cal H}_{\kappa}$ carrying the UIR $(p,q)$ differs from the
highest weight state.

We can now obtain the replacement for the previous
eqn.$(\ref{89})$.  It is unavoidably somewhat more complicated.  
Using eqn.$(\ref{113})$ and with $\kappa =uv(x+iy)$, we have:
     \begin{eqnarray*}
     |\underline{z}^{(0)}(u), \underline{w}^{(0)}(v,x,y)
     \rangle = e^{-\frac{1}{2}(u^2+v^2)}\sum\limits
     ^{\infty}_{p,q=0}\;u^p\;v^q\left\{_0F_1(2k;|\kappa|^2)
     /(p+q+1)!\right\}^{1/2}\times
     \end{eqnarray*}
     \[{\cal N}^{\prime}(p,q; |\kappa|/uv)\;|p,q;  
     \kappa/uv\rangle_{\kappa} ,\]
     \begin{eqnarray*}
     {\cal N}^{\prime}(p,q; |\kappa|/uv)\;|p,q;  
     \kappa/uv\rangle_{\kappa} =
     \sum\limits^{I_0}_{I=|M_0|}\;\sum\limits^{I}
     _{M=M_0}\;\frac{(-1)^{I_0-M}}{(M-M_0)!}\;
     (\kappa/uv)^{I_0-M}\times
    \end{eqnarray*}
     \begin{eqnarray*}
     \left(1-\frac{|\kappa|^2}{u^2 v^2}\right)^{\frac{1}{2}
     (M-M_0)}&&\{(2I+1)(I-M_0)! (I+M)!/(I+M_0)!
      (I-M)!\}^{1/2}\nonumber\\
      &&  |p,q;I,M,\frac{1}{3}(p-q)\rangle_{\kappa},
     \end{eqnarray*}
     \begin{eqnarray}
     {\cal N}^{\prime}(p,q;|\kappa|/uv) &=&
     \left\{\sum\limits^{I_0}_{I=|M_0|}\;
     \sum\limits^I_{M=M_0}\;\frac{(2I+1)(I-M_0)!(I+M)!}
     {(I+M_0)!(I-M)!(M-M_0)!^2}\right.\nonumber\\
     &&\left.\right.\nonumber\\
     &&\left.(|\kappa|/uv)^{2(I_0-M)}\left(1-\frac{|\kappa|^2}
     {u^2v^2}\right)^{M-M_0}\right\}^{1/2},
     \label{114}
     \end{eqnarray}
which, as shown in the Appendix, can be compactly written as
\begin{eqnarray}
{\cal N}^{\prime}(p,q;|\kappa|/uv)& =&\left\{(I_0-|M_0|+1)
(I_0+|M_0| +1) \right.\nonumber\\
&&\left.  _2F_1(-(I_0-|M_0|), -(I_0+|M_0|), 2, 1- |\kappa|^2/(u^2 v^2))
\right\}^{1/2} 
\label{131}
\end{eqnarray} 

\noindent
The normalisation factor ${\cal N}^{\prime}(p,q;|\kappa|/uv)$ has 
been defined so as to make the vector $|p,q;\kappa/uv \rangle
_{\kappa}$ have unit norm; this vector lies in the subspace of
${\cal H}_{\kappa}$ carrying the (single occurrence of the) UIR $(p,q)$ 
in ${\cal D}_{\kappa}$.  Now we apply ${\cal U}(A)$ to both sides of
eqn.$(\ref{114})$ and get the replacements for eqns.$(\ref{90}, \ref{91})$:
     \[A\in\;SU(3),\;\underline{z}=A\;\underline{z}^{(0)}(u),\;
      \underline{w}=A^*\underline{w}^{(0)}(v,x,y),
      \underline{z}^T\underline{w}=\kappa :\]
     \begin{eqnarray}
     |\underline{z},\underline{w}\rangle &=& e^{-\frac{1}{2}
     (u^2+v^2)}\sum\limits^{\infty}_{p,q=0} u^p\;v^q\nonumber\\
     &&\left\{_0F_1(2k;|\kappa|^2)/(p+q+1)!\right\}^{1/2}
      {\cal N}^{\prime}(p,q;|\kappa|/uv)\times\nonumber\\
      &&|p,q;\kappa/uv; A\rangle_{\kappa} ,\nonumber\\ 
      |p,q;\kappa/uv; A\rangle_{\kappa}&=&
      {\cal U}(A)\;|p,q;\kappa/uv \rangle_{\kappa}.
     \label{115}
     \end{eqnarray}

\noindent
We see that for $(\underline{z},\underline{w})\in{\cal C}^6$ with 
given $u,v,\kappa$, corresponding to Class (d) in eqn.$(\ref{79})$, the H-W
SCS $|\underline{z},\underline{w}>$ is expressible in terms
of a sequence of $SU(3)$ GCS, all contained in ${\cal H}_{\kappa}$.
The $SU(3)$ GCS within the UIR $(p,q)$ use 
$|p,q;\kappa/uv\rangle_{\kappa}$ as the fiducial vector, 
and this is very different from the highest weight vector.
For this family of $SU(3)$ GCS we have in place of eqn.(4.32):
     \begin{eqnarray}
     \int\limits_{SU(3)}dA\;|p,q;\kappa/uv; A\rangle_{\kappa}\;
      _{\kappa}\langle p^{\prime},q^{\prime}; \kappa/uv;
      A|=\delta_{p^{\prime}p}\delta_{q^{\prime}q}\;
      \frac{P^{(p,q;\kappa)}}{d(p,q)} \;,
      \label{116}
      \end{eqnarray}

\noindent
where $P^{(p,q;\kappa)}$ is the projection operator onto
the subspace of ${\cal H}_{\kappa}$ carrying the UIR $(p,q)$.
This follows from  Schur lemma for $SU(3)$ UIR's, and
the fact that ${\cal D}_{\kappa}$ is multiplicity-free.

With these replacements for eqns.$(\ref{89},\ref{90}, \ref{91}, \ref{95})$ in 
hand, we can study the analogue of the operator $A(f_0)$ in eqn.$(\ref{93})$.  
We begin with the general definition $(\ref{86})$ of $A(f)$ and choose
     \begin{eqnarray}
     f(u,v,x,y)&=&f_0(u,v) \delta^{(2)}(x+iy-\kappa/uv)\nonumber\\
     &=&f_0(u,v) \delta(x-\mbox{Re}\;\kappa/uv) \delta/y
     -\mbox{Im}\;\kappa/uv ).
    \end{eqnarray}

\noindent
This achieves the restriction to ${\cal H}_{\kappa}$.
We then define
     \begin{eqnarray}
     A(f_0)&=& \int\prod\limits^3_{j=1}
     \left(\frac{d^2z_j}{\pi}\;\frac{d^2w_j}{\pi}\right)
     f_0(u,v) \delta^{(2)} (x+iy-\kappa/uv)
     |\underline{z},\underline{w}\rangle\langle\underline{z},
     \underline{w}|\nonumber\\
     &=&\frac{2}{\pi}\int\limits^{\infty}_0 u^5du\int\limits
      ^{\infty}_0 v^5 dv f_0(u,v) \theta(uv-|\kappa|)
     \left(1-\frac{|\kappa|^2}{u^2v^2}\right)\times\nonumber\\
     &&\int\limits_{SU(3)}dA\;{\cal U}(A)|\underline{z}^{(0)}(u),
     \underline{w}^{(0)}(v,x,y)\rangle\langle\underline{z}^{(0)}
     (u),\underline{w}^{(0)}(v,x,y)|{\cal U}(A)^{-1},
     \end{eqnarray}

\noindent
it being understood in the last expression that $x+iy=\kappa/uv$.
We can now use eqns.$(\ref{115}, \ref{116})$ here and get the final result
replacing eqn.$(\ref{96})$:
     \begin{eqnarray*}
     A(f_0)&=&\sum\limits^{\infty}_{p,q=0} C(f_0;p,q;\kappa)\;
     P^{(p,q;\kappa)},\nonumber\\
     C(f_0;p,q;\kappa)&=&\frac{2}{\pi}
     \left\{_0F_1(2k;|\kappa|^2)/(p+q+1)! d(p,q)\right\}
     \int\limits^{\infty}_{0} u^5 du \int\limits^{\infty}_0
     v^5 dv\;f_0(u,v)\times\nonumber\\
     \end{eqnarray*}  
     \begin{eqnarray}
     \theta(uv-|\kappa|)\left(|-\frac{|\kappa|^2}{u^2v^2}\right)
     e^{-(u^2+v^2)} u^{2p} v^{2q}
     \{{\cal N}^{\prime}(p,q;|\kappa|/uv)\}^2.
     \label{119}
     \end{eqnarray}

\noindent
The freedom remaining in the choice of $f_0(u,v)$ displays
the overcompleteness, within ${\cal H}_{\kappa}$, of the H-W
SCS belonging to ${\cal H}_{\kappa}$.  To limit ourselves to an 
exact ``$SU(3)$- worth'' of these states, we choose $f_0(u,v)$ to be the 
product of two delta functions.  Then we get a generalisation
of eqn.$(\ref{97})$:
     \begin{eqnarray}
     f_0(u,v)&=& \delta(u-u_0) \delta(v-v_0)\;,\;u_0v_0>|\kappa|:
     \nonumber\\
      &&\nonumber\\
     A(f_0)&=& \int\;\prod\limits^3_{j=1}\left(\frac{d^2z_j}
     {\pi}\;\frac{d^2w_j}{\pi}\right)
     \delta(u-u_0) \delta(v-v_0) \delta^{(2)}
     \left(x+iy -\frac{\kappa}{uv}\right) |\underline{z},
     \underline{w}\rangle\langle\underline{z},\underline{w}|
     \nonumber\\
     &&\nonumber\\
     &=&\frac{2}{\pi}\cdot e^{-\left(u_0^2+v_0^2\right)}
     \sum\limits^{\infty}_{p,q=0}\left\{_0F_1(2k;|\kappa|^2)/
     (p+q+1)!d(p,q)\right\}\times\nonumber\\
     &&\nonumber\\
     &&u_0^{2p+5} v_0^{2q+5} \left(1-\frac{|\kappa|^2}
     {u^2_0 v^2_0}\right) \left\{{\cal N}^{\prime}
     (p,q;|\kappa|/u_0v_0 )\right\}^2\;
      P^{(p,q;\kappa)}.
     \label{120}
     \end{eqnarray}

In this manner all the results found in the preceeding Section
for the subspace ${\cal H}_0\subset {\cal H}$, the null space
of $K_-$, generalise to a general eigenspace ${\cal H}_{\kappa}
\subset {\cal H}$ of $K_-$.  Here again, limiting oneself to
an exact ``$SU(3)$-worth'' of H-W SCS does give us a total set
of vectors, but they do not obey the Klauder resolution of the 
identity within ${\cal H}_{\kappa}$       

\section{H-W SCS of Class(e) and their $SU(3)$ content}

In the listing of $SU(3)$ orbits of H-W SCS given in eqn.$(\ref{79})$,
it was pointed out that only Classes (d) and (e) involve all
six oscillators of the Schwinger $SU(3)$ construction in a
nontrivial manner.  Furthermore, of these, only the former are 
generic.  As we have seen, Class (d) orbits form a four-parameter
continuous family, each orbit being of dimension eight.  In
contrast, Class (e) orbits are a three parameter family, with
each orbit of dimension five.  Another characteristic is that
each H-W SCS $|\underline{z},\underline{w}>$ in Class (d) is
such that the complex three-vectors $\underline{z}^*$ and 
$\underline{w}$ are linearly independent; on the other hand,
if $|\underline{z},\underline{w}>$ is in Class (e), then
$\underline{w}$ is a (complex) multiple of $\underline{z}^*$.

In Sections IV and V we have analysed in detail the 
$SU(3)$ structure and representation content of H-W SCS
on all Class (d) orbits, for $\underline{z}^T\underline{w}=0$
and $\underline{z}^T\underline{w} = \kappa\neq 0$ respectively.
Now we turn to a similar analysis of the Class (e) orbits\cite{30}.  
There is however a difficulty in handling this case by starting 
with the Klauder resolution of the identity, eqns.$(\ref{75})$, and then 
modifying the integrand by inserting some function of the
$SU(3)$ invariants with the aim of restricting the integration to 
a chosen subset of orbits.  We are unable to use the methods
of Sections IV and V here.  The reason is that in terms of the
$SU(3)$ invariant parameters $u,v,x,y$ in eqn.$(\ref{78})$, Class (e)
corresponds to $x^2+y^2=1$; while in the volume element
$(\ref{84})$ on the H-W group there is an explicit factor $(1-x^2-y^2)$.
For this reason, we handle Class (e) orbits more directly,
guided however by the results in Class (d).

A convenient representative point on a general Class (e)
orbit is given by the pair of complex three-vectors
     \begin{eqnarray}
     \underline{z}^{(0)}(u)&=&u(0,0,1)^T\;,\;u>0,\nonumber\\
      \underline{w}^{(0)}(ve^{i\alpha}) &=& ve^{i\alpha}(0,0,1)^T
      \;,\; v>0,\;,\;0\leq \alpha < 2\pi ,\nonumber\\
      &&\underline{z}^{(0)}(u)^T \underline{w}^{(0)} (ve^{i\alpha})
      = u v e^{i\alpha}.
      \end{eqnarray}

\noindent
(As mentioned earlier in Section IV, the reason for choosing this 
configuration is that the corresponding
stability group is the $SU(2)$ subgroup acting on dimensions 1
and 2 in the defining representation $(1,0)$, and it is just this
subgroup that is involved in the canonical basis vectors
$|p,q;IMY\rangle$ in a general $SU(3)$ UIR $(p,q)$).  Acting
with a general $A\in\;SU(3)$, we reach a general point
$(\underline{z},\underline{w})$ on the orbit given by
     \begin{eqnarray}
     \underline{z}&=& A\;\underline{z}^{(0)}(u),\nonumber\\
     \underline{w}&=& A^*\underline{w}^{(0)}(ve^{i\alpha}) =
     \left(\frac{ve^{i\alpha}}{u}\right)\;\underline{z}^*,
     \nonumber\\
     \underline{z}^T\underline{w}&=& u v e^{i\alpha} = e^{i\alpha}
     \left(\underline{z}^{\dag}\underline{z}\;\underline{w}
     ^{\dag}\underline{w}\right)^{1/2}.
     \end{eqnarray}

\noindent
The H-W SCS $|\underline{z}^{(0)}(u), \underline{w}^{(0)}
(ve^{i\alpha})\rangle$ is of course given by
        \begin{eqnarray}
         |\underline{z}^{(0)}(u), \underline{w}^{(0)}
         (ve^{i\alpha})\rangle = 
         e^{-\frac{1}{2}(u^2+v^2)+u \hat{a}^{\dag}_3 +
         ve^{i\alpha} \hat{b}_3^{\dag}}\;|\underline{0},
         \underline{0}>.
         \end{eqnarray}

\noindent
We can expand this in the orthonormal basis $|p,q;IMY; m>$ for
${\cal H }$, recognising that the only states that appear
have $I=M=0, Y=\frac{2}{3}(q-p)$ for various $(p,q)$.
We need the results (I.A.6, I.A.7):
     \begin{mathletters}
     \begin{eqnarray}
     |p,q; 0,0,\frac{2}{3}(q-p);k\rangle &=&
     p! q! \{(p+1)(q+1)/(p+q+1)!\}^{1/2}\times\nonumber\\
     \sum\limits^{(p,q)_<}_{n=0}&&\frac{(-1)^n}{(n+1)!}\;
     \frac{\left(\hat{a}_{\alpha}^{\dag}\hat{b}^{\dag}_{\alpha}
     \right)^n}{n!}\;\frac{\left(\hat{a}^{\dag}_3\right)^{p-n}}
     {(p-n)!}\;\frac{\left(\hat{b}^{\dag}_3\right)^{q-n}}
     {(q-n)!}\;|\underline{0},\underline{0}\rangle ,\nonumber\\
     \hat{a}_{\alpha}^{\dag} \hat{b}^{\dag}_{\alpha}&=&
     \hat{a}_1^{\dag} \hat{b}_1^{\dag} + \hat{a}^{\dag}_2
     \hat{b}^{\dag}_2 ;\\
     &&\nonumber\\
     |p,q;0,0,\frac{2}{3}(q-p);m\rangle&=&
     \{(2k-1)!/(m-k)! (m+k-1)!\}^{1/2} \cdot
     \left(\underline{\hat{a}}^{\dag}\cdot
     \underline{\hat{b}}^{\dag}\right)^{m-k}\times\nonumber\\
     &&|p,q;0,0,\frac{2}{3}(q-p); k\rangle .
      \end{eqnarray}
      \label{124}
      \end{mathletters}

\noindent
We can now easily compute the desired overlap:
     \begin{eqnarray*}
     \langle p,q;0,0,\frac{2}{3}(q-p); m| \underline{z}^{(0)}(u),
      \underline{w}^{(0)} (ve^{i\alpha})\rangle =
     \{(2k-1)!/(m-k)!(m+k-1)!\}^{1/2}\times
     \end{eqnarray*}
     \[(uve^{i\alpha})^{m-k}\langle p,q;0,0,\frac{2}{3}
      (q-p);k|\underline{z}^{(0)}(u),\underline{w}^{(0)}
      (ve^{i\alpha})\rangle\]
      \[=\{(2k-1)!/(m-k)!(m+k-1)!\}^{1/2}
      (uve^{i\alpha})^{m-k}\cdot p!q!
      \{(p+1)(q+1)/(p+q+1)!\}^{1/2}\times\]
     \begin{eqnarray*}
      \langle\underline{0},\underline{0}|
      \frac{\hat{a}_3\;^p}{p!}\;\frac{\hat{b}_3\;^q}{q!}\;
      |\underline{z}^{(0)}(u), \underline{w}^{(0)}(ve^{i\alpha})
      \rangle
      \end{eqnarray*}
      \[\{(p+1)(q+1)(2k-1)!/(p+q+1)!(m-k)!(m+k-1)!\}^{1/2}\;
      (uve^{i\alpha})^{m-k}\times\]
      \begin{eqnarray}
      e^{-\frac{1}{2}(u^2+v^2)} \;u^p(v e^{i\alpha})^q .
      \end{eqnarray}

\noindent
In the second step here, when using the expansion
$(\ref{124}a)$, only the term $n=0$ contributes.  We therefore have
the expansion of the representative Class (e) H-W SCS in the
$SU(3) \times Sp(2,R)$ basis:
     \begin{eqnarray*}
     |\underline{z}^{(0)}(u),\underline{w}^{(0)}
     (ve^{i\alpha})\rangle &=& e^{-\frac{1}{2}(u^2+v^2)}
    \sum\limits^{\infty}_{p,q=0} 
     \{(p+1)(q+1)/(p+q+1)!\}^{1/2} \;u^p(ve^{i\alpha})^q\times
    \nonumber\\
    &&\nonumber\\
    &&\sum\limits^{\infty}_{m=k}\{(2k-1)!/(m-k)!(m+k-1)!\}^{1/2}
    (uve^{i\alpha})^{m-k}|p,q;0,0,\frac{2}{3}(q-p);m\rangle
    \end{eqnarray*}
    \begin{eqnarray}
    &=&e^{-\frac{1}{2}(u^2+v^2)}\sum\limits^{\infty}_{p,q=0}
    \{(p+1)(q+1/(p+q+1)!\}^{1/2}\;
    u^p(ve^{i\alpha})^q\left\{_0F_1(2k;u^2v^2)\right\}^{1/2}\times
     \nonumber\\
     &&\nonumber\\
     &&|p,q;0,0,\frac{2}{3} (q-p)\rangle_{uve^{i\alpha}}.
     \label{126}
     \end{eqnarray}

\noindent
As we would expect, this expansion involves just the $K_-$ 
eigenstate defined in eqn.(5.5), namely the $I=M=0, 
Y=\frac{2}{3}(q-p)$ member of the orthonormal basis 
$\{|p,q;IMY\rangle_{uve^{i\alpha}}\}$ for ${\cal H}_{uve^{i\alpha}}$.
As in the case of the $SU(3)$ SCS, where the fiducial vector
within the UIR $(p,q)$ is the single highest weight vector
$|p,q;\frac{1}{2}(p+q),\frac{1}{2}(p+q),\frac{1}{3}(p-q)\rangle$,
here too a single vector of the canonical basis appears
as fiducial vector, but it is of course not the highest 
weight state.

Now within each UIR $(p,q)$ contained in the UR
${\cal D}_{uve^{i\alpha}}$ on ${\cal H}_{uve^{i\alpha}}$,
we define the family of $SU(3)$ GCS:
     \begin{eqnarray}
     A\in\;SU(3) : |p,q;0,0,\frac{2}{3}(q-p); 
     A\rangle_{uve^{i\alpha}} ={\cal U}(A) |p,q;0,0,\frac{2}{3}
     (q-p)\rangle_{uve^{i\alpha}}
     \label{127}
     \end{eqnarray}

\noindent
Then applying ${\cal U}(A)$ to both sides of eqn.$(\ref{126})$ we have
the general connection between Class (e) H-W SCS and the
$SU(3)$ GCS $(\ref{127})$:
     \begin{eqnarray*}
     |A \underline{z}^{(0)}(u),\;A^*\underline{w}^{(0)}
     (ve^{i\alpha})\rangle =
     e^{-\frac{1}{2}(u^2+v^2)}\sum\limits^{\infty}
     _{p,q=0}\{(p+1)(q+1)/(p+q+1)!\}^{1/2}\times
     \end{eqnarray*}
     \begin{eqnarray}
     u^p(ve^{i\alpha})^q\left\{_0F_1(2k;u^2v^2)\right\}^{1/2}
     |p,q;0,0,\frac{2}{3}(q-p); A\rangle_{uve^{i\alpha}}
     \label{128}
     \end{eqnarray}

\noindent
We recognise that eqns.$(\ref{126}, \ref{127},\ref{128})$ are replacements 
for eqns. $(\ref{89}, \ref{90},\ref{91})$ and eqns.$(\ref{114}, \ref{115})$ 
of Class (d).

Keeping $uve^{i\alpha}$ fixed, the $SU(3)$ GCS $(\ref{127})$ all
belong to ${\cal H}_{uve^{i\alpha}}$, and from Schur lemma 
they obey the analogues to eqns.$(\ref{95}, \ref{116})$:
     \begin{eqnarray}
     \int\limits_{SU(3)}\;dA |p,q;0,0,\frac{2}{3}(q-p);
     A\rangle_{uve^{i\alpha}}\;_{uve^{i\alpha}}\langle
     p^{\prime},q^{\prime}; 0,0,\frac{2}{3}(q^{\prime}-
     p^{\prime}); A|=\delta_{p^{\prime}p}
     \delta_{q^{\prime}q}\;
     \frac{P^{(p,q;uve^{i\alpha})}}{d(p,q)},
     \end{eqnarray}

\noindent
Here of course we exploit the multiplicity-free reduction of
${\cal D}_{uve^{i\alpha}}$.  It follows that for the H-W SCS $(\ref{128})$
we have:
     \begin{eqnarray*}
      \int\limits_{SU(3)}\;dA |A\underline{z}^{(0)}(u), A^*
      \underline{w}^{(0)}(ve^{i\alpha})\rangle\langle
      A\underline{z}^{(0)}(u), A^*\underline{w}^{(0)}
      (ve^{i\alpha})|=
     \end{eqnarray*}
      \begin{eqnarray}
      2e^{-(u^2+v^2)}\sum\limits^{\infty}_{p,q=0}\;u^{2p}v^{2q}\;
      _0F_1(2k;u^2v^2)\;P^{(p,q;uve^{i\alpha})}/
      (p+q+2)!
      \label{130}
      \end{eqnarray}

\noindent
The integration over $SU(3)$ here is in effect only over the
five-dimensional coset space $SU(3)/SU(2)$, in contrast to 
eqns.$(\ref{95}, \ref{116})$ in Class (d).

If we write $\kappa=uve^{i\alpha}$ and allow $u$ and $v$ to
vary reciprocally, and also keep $\alpha$ fixed so that             
$\kappa$ stays fixed, we never leave the subspace 
${\cal H}_{uve^{i\alpha}}$ and the projection operators
$P^{(p,q;uve^{i\alpha})}$.  Therefore we can multiply
both sides of eqn.$(\ref{130})$ by any function
     \begin{eqnarray}
     f(u,v)=f_0(u) \;\delta(uv-|\kappa|),
     \end{eqnarray}

\noindent
and integrate over both $u$ and $v$ to get results similar to
eqns.$(\ref{96}, \ref{119})$.  Here $f_0(u)$ is free.  This then
shows that for each fixed $\kappa$, the Class (e) H-W SCS
$|\underline{z},\underline{w}>$ with $\underline{z}^T
\underline{w}=\kappa$ are overcomplete in ${\cal H}_{\kappa}$

\section{Concluding remarks}
To conclude, we have given  a unified analysis of the interconnections between 
the Heisenberg-Weyl standard coherent states and the standard coherent states 
as well as certain generalised coherent states of $SU(3)$. The specific family
of $SU(3)$ coherent states to be used is dependent on the type of orbit of 
the H-W SCS belong to. This situation is describable in detail as follows.  
In terms of the $SU(3)$ invariant parameters $x$ and $y$, at $x=y=0$ we have 
those generic Class(d) orbits which lie entirely within the subspace 
${\cal H}_0$. For these H-W SCS, the $SU(3)$ harmonic analysis involves 
precisely the $SU(3)$ SCS within each UIR.  For $0<x^2+y^2<1$ we deal with 
the subspaces ${\cal H}_{\kappa}\subset {\cal H}$ which generalise 
${\cal H}_0$; the corresponding orbits consist of H-W SCS whose $SU(3)$ content
brings in the $SU(3)$ GCS studied in Section V.  The fiducial vectors here 
are rather complicated, at any rate in the canonical basis for 
$SU(3)$ UIR's.  In the limit $x^2+y^2=1$, we have the Class (e) orbits.
These H-W SCS involve yet another family of $SU(3)$ GCS, though now the 
fiducial vectors are the unique $SU(2)$ scalar states within each $SU(3)$ UIR,
and their properties are studied in Secion VI. 
In this entire development the group $Sp(2,R)$ plays 
a particularly helpful role and so does the  Schur lemma wherever it is 
available. Indeed we have used this lemma for UIR's of the H-W group 
wherever possible, and after modifications of the completeness identity used 
it for UIR's of $SU(3)$. This systematic use of Schur lemma makes several 
computations much easier than otherwise.  It must be emphasised 
that all the Heisenberg  standard coherent states have been included in our 
study in the spirit of $SU(3)$ harmonic analysis, so that there is a
satisfactory completeness in our analysis. The significant property of the 
discrete series UIR's of $Sp(2,R)$, which we have exploited, is worth
mention. It is that while the spectrum of the compact generator $J_0$ depends 
on $k$, hence on the UIR, the `spectrum' of the non hermitian lowering
operator $K_-$ is the entire complex plane, thus being UIR independent. The 
calculations in Section V clearly show the importance of these facts.  

\newpage
\def\theequation{A.\arabic{equation}}
\appendix{\bf{Appendix}}
\setcounter{equation}{0}

We outline  here the steps involved in going from $(\ref{114})$ to 
$(\ref{131})$. Eqn. $(\ref{114})$. 
\begin{eqnarray}
     {\cal N}^{\prime}(p,q;|\kappa|/uv) &=&
     \left\{\sum\limits^{I_0}_{I=|M_0|}\;
     \sum\limits^I_{M=M_0}\;\frac{(2I+1)(I-M_0)!(I+M)!}
     {(I+M_0)!(I-M)!(M-M_0)!^2}\right.\nonumber\\
     &&\left.\right.\nonumber\\
     &&\left.(|\kappa|/uv)^{2(I_0-M)}\left(1-\frac{|\kappa|^2}
     {u^2v^2}\right)^{M-M_0}\right\}^{1/2},
    \label{132}
    \end{eqnarray}
can be written in terms of the Jacobi polynomials 
\begin{equation}
P_{n}^{(\alpha, \beta)}(x)\equiv
\frac{\Gamma(\alpha+n+1)}{n!\Gamma(\alpha+\beta+n+1)}
\sum_{m=0}^{n} {n \choose m} \frac{\Gamma(\alpha +\beta + n+m+1)}
{2^m \Gamma(\alpha +m +1)}(x-1)^m
\end{equation} 
as 
\begin{equation}
{\cal N}^{\prime}(p,q;|\kappa|/uv)= 
\left\{\sum_{I=|M_0|}^{I_0} (2I+1)\left(\frac{|\kappa|}{uv}
\right)^{2(I_0-M_0)} P_{I-M_0}^{(0,2M_0)}\left(\frac{2u^2v^2}{|\kappa|^2}-1
\right) \right\}^{1/2}
\end{equation}
Using the fact that $P_n^{(\alpha,\beta)}$ can also be written as  
\begin{equation}
P_{n}^{(\alpha, \beta)}(x) =
\frac{1}{2^n}
\sum_{m=0}^{n} {n+\alpha \choose m} {n+\beta \choose n-m}(x-1)^{n-m}(x+1)^m
\end{equation}
one can show that 
\begin{equation}
x^{M_0} P_{I-M_0}^{(0,2M_0)}(2x-1) = x^{-M_0}P_{I+M_0}^{(0,-2M_0)}(2x-1)
\end{equation}
which implies that ${\cal N}^{\prime}$ depends on $M_0$ only through its 
magnitude $|M_0|$. Replacing $M_0$ in the rhs of $(\ref{132})$ by $|M_0|$ 
and rewriting it as a polynomial in $(1-|\kappa |^2/u^2 v^2)$, we obtain
\begin{equation}
{\cal N}^{\prime}(p,q;|\kappa|/uv)= 
\left\{\sum_{k=0}^{I_0-|M_0|} a_k \left(1-\frac{|\kappa|^2}{u^2v^2}\right)^k
 \right\}^{1/2}, 
\label{133}
\end{equation}
where 
\begin{equation}
a_k = \sum_{M=0}^{k}\sum_{I=M}^{I_0-|M_0|}(2I+2|M_0|+1)
{ I+2|M_0|+M \choose M} { I \choose M} {I-|M_0|-M \choose k-M} (-1)^{k-M}.
\end{equation}
which, after some rearrangement, can be written as 
\begin{eqnarray}
a_k=&&\sum_{M=0}^{k}(-1)^{M}{I_0-|M_0|+M-k\choose M}\cdot\nonumber\\
&&\sum_{I=0}^{I_0-|M_0|-k+M}(2I+2k+2|M_0|-2M +1)
{ I+k-M \choose I} {I+2k+2|M_0|-2M \choose k-M}.
\end{eqnarray}
Using the identities 
\begin{eqnarray}
\sum_{I=0}^{I_0-|M_0|-k+M}&&(2I+2k+2|M_0|-2M +1)
{ I+k-M \choose I} {I+2k+2|M_0|-2M \choose k-M}\nonumber\\
&& = (I_0-|M_0|+1)
{I_0 +|M_0|+k-M+1 \choose I_0 +|M_0|} {I_0-|M_0|\choose k-M}
\end{eqnarray}
and 
\begin{eqnarray}
\sum_{M=0}^{k}&&(-1)^{M} {I_0-|M_0|+M-k\choose M}
{I_0 +|M_0|+k-M+1 \choose I_0 +|M_0|} {I_0-|M_0|\choose k-M}\nonumber\\
&&={I_0 + |M_0| +1 \choose k+1}{I_0-|M_0| \choose k}
\end{eqnarray}
we  obtain
\begin{equation}
a_k = (k+1){I_0 + |M_0| +1 \choose k+1}{I_0-|M_0|+1 \choose k+1}
\end{equation}
Substituting this in $(\ref{133})$ we finally obtain the result $(\ref{131})$.


\begin{references}
\bibitem{11}
These are given in eqn.(4.11) below.
\bibitem{14}
The use of a common symbol $\psi$ for elements in ${\cal H}_0$ and 
${\cal H}^{({\rm ind})}$ should cause no confusion.
\bibitem{20}
The totality property of a set of vectors means that any vector orthogonal to 
all of them necessarily vanishes.
\bibitem{22}
Here we use an important result quoted in the book of A. M. Perelomov, 
ref 5 above, appendix B, involving the associated Laguerre polynomials.
\end{references}
\end{document}